\documentclass[a4paper,fleqn, times]{cas-dc}

\usepackage[numbers]{natbib}
\usepackage{subfigure}
\usepackage{color}
\usepackage{tcolorbox}
\usepackage{multirow}
\usepackage{graphicx}
\usepackage{hyperref}
\usepackage[switch]{lineno}

\def\tsc#1{\csdef{#1}{\textsc{\lowercase{#1}}\xspace}}
\tsc{WGM}
\tsc{QE}

\begin{document}
\let\WriteBookmarks\relax
\def\floatpagepagefraction{1}
\def\textpagefraction{.001}

\shorttitle{Advancing Autonomous Driving System Testing: Demands, Challenges, and Future Directions}

\shortauthors{Liao et al.}   

\title [mode = title]{Advancing Autonomous Driving System Testing: Demands, Challenges, and Future Directions}

\author[1]{Yihan Liao}[orcid=0000-0002-8002-9190]
\ead{yihanliao4-c@my.cityu.edu.hk}

\author[1]{Jingyu Zhang}[orcid=0000-0001-6043-4239]
\cormark[1]
\ead{jzhang2297-c@my.cityu.edu.hk}

\author[1]{Jacky Keung}[orcid=0000-0002-3803-9600]
\ead{jacky.keung@cityu.edu.hk}

\author[2]{Yan Xiao}[orcid=0000-0002-2563-083X]
\ead{xiaoyan.hhu@gmail.com}

\author[3]{Yurou Dai}[orcid=0000-0002-2553-6915]
\ead{yrd224@lehigh.edu}
            
\affiliation[1]{organization={City University of Hong Kong},
            city={Hong Kong},
            country={China}}

\affiliation[2]{organization={Shenzhen Campus of Sun Yat-sen University}, 
            city={Shenzhen},
            country={China}}
            
\affiliation[3]{organization={Lehigh University}, 
            city={Bethlehem},
            country={USA}}

\cortext[1]{Corresponding author}

\begin{abstract}
\noindent \textbf{Context:} 
Autonomous driving systems (ADSs) promise enhanced transportation efficiency but face critical challenges in ensuring reliability across complex driving environments. Effective testing is essential to validate ADS performance and mitigate real-world risks.

\noindent \textbf{Objective:}
This study investigates current ADS testing practices for both modular and end-to-end systems, identifies key demands (needs required by practitioners and researchers), and examines gaps between research and real-world demands. We review critical testing techniques and extend to involve Vehicle-to-Everything (V2X) communication and Foundation Models (FMs), including large language models and vision foundation models, in enhancing ADS testing performance. We provide literature reviews and outline future directions for each demand of industry practitioners and academic researchers.

\noindent \textbf{Methods:}
A large-scale survey was conducted with 100 participants, including industry practitioners and academic researchers. We first discuss survey questions with professionals, distribute them to industry practitioners and academic researchers, and conduct follow-ups. Quantitative and qualitative analyses uncover key trends and challenges.

\noindent \textbf{Results:}
Findings reveal that existing ADS testing techniques struggle to evaluate real-world performance comprehensively, including the diversity of corner cases, the gap between simulation and real-world testing, the lack of comprehensive testing criteria, potential attacks, practical deployment for V2X, and the computational costs for FMs. By analyzing participants' responses and 105 relevant papers, we summarize the future research directions.

\noindent \textbf{Conclusion:}
Our study highlights critical research gaps in ADS testing and underscores the demands of industry practitioners and academic researchers. We provide future directions for ADS: comprehensive testing criteria, cross-model collaboration in V2X, enhancing cross-modality (e.g., text and image) adaptation in FM testing, and scalable ADS validation frameworks. These insights contribute to advancing software engineering practices for ADS development, ensuring safer and more reliable autonomous systems.

\end{abstract}



\maketitle

\section{Introduction}
The rapid advancements in autonomous driving systems (ADSs) have revolutionized modern transportation, but safety concerns and testing challenges remain critical barriers to widespread adoption \cite{tesla2020, waymo2021, apollo2021}. 
Testing autonomous driving systems requires addressing complex challenges, including corner case diversity, real-world simulation fidelity, and the integration of emerging technologies like Vehicle-to-Everything (V2X) communication. For instance, leading industry players such as Tesla launched the Full Self-Driving (FSD) Beta program in 2020, allowing experienced drivers to test the latest ADS capabilities \cite{tesla2020}. Waymo tested Level 4 \cite{on2021taxonomy} autonomous driving in multi-city environments and launched the autonomous driving taxi service in 2018, which is now employed more than a hundred thousand times per week \cite{waymo2021}. Apollo Go is subject to Baidu's autonomous driving travel service, which started in 2020 and has already provided nearly nine hundred thousand rides \cite{apollo2021}. However, despite these advancements, the safety and reliability of ADSs remain a major concern, with reports of nearly 400 crashes, particularly involving Tesla's ADS \cite{report2022}. Therefore, rigorous testing methodologies are crucial to ensuring the robustness of ADSs, a topic extensively explored in the software engineering community.

In recent years, the software engineering community has presented many testing insights and approaches, one of the prominent fields is test case generation, which aims to create diverse and challenging driving scenarios for ADS validation \cite{tian2022generating, ding2023survey, lu2023test, huai2023sceno, guo2024sovar, song2024empirically, zhang2024chatscene}. Besides, to reduce the costs of ADS real-world testing, simulation platforms have emerged as an effective alternative \cite{dosovitskiy2017carla, apollo2021, awsim2022, beamng2022} by generating a large number of driving test scenarios. V2X communication is another promising area of research that enhances information exchange between vehicles and their surroundings, allowing wireless communication with vehicles, infrastructure, and pedestrians, improving safety and efficiency. Its integration adds complexity to testing, as cooperative ADSs require validation at both the individual vehicle and multi-agent/system-wide levels. Existing testing frameworks mainly focus on single-agent performance, neglecting V2X interactions \cite{xu2023v2v4real}. This survey examines V2X testing practices, identifying gaps between research and real-world implementation and providing insights into future directions for ensuring V2X-enabled ADSs' robustness and reliability \cite{sedar2021standards, martinez2021security, yang2022edge}.

Additionally, emerging Foundation Models (FMs), such as Large Language Models (LLMs) and Vision Foundation Models (VFMs), are being integrated into ADS testing to improve methodologies \cite{zhang2023automated, guo2024sovar}. LLMs can automate scenario generation, create adversarial corner cases, and provide natural language explanations for test results \cite{lu2024multimodal, wei2024editable, zhang2024chatscene, lu2024diavio}. VFMs enhance perception by synthesizing realistic environments and identifying critical failure cases \cite{caron2021emerging, kappeler2024few}. Current research mainly explores theoretical applications, with few surveys on the practical integration of FMs into ADS testing frameworks. This gap highlights the need to explore how FMs can improve testing methodologies. Thus, we aim to investigate FMs' role in ADS testing, identify challenges, and propose future directions to enhance ADS robustness and reliability.

With the rapid development of ADS testing techniques, many survey papers have summarized and compared them \cite{lou2022testing, tang2023survey}. The most related paper to our work is from Lou \textit{et~al.} \cite{lou2022testing}, which conducted a survey and literature review on ADS testing. However, compared to their work, we have (1) covered a broader scope by incorporating over three times the number of survey questions as their study (31 questions) with a similar number of responses and analyzed the results separately for academic and industry participants, (2) expanded the discussion to include testing criteria, potential attacks, V2X collaboration, FMs, etc. Consequently, a gap exists in understanding how emerging research aligns with the current ADS testing demands of industry practitioners and academic researchers, particularly for Level-3 to Level-5 ADSs as defined by the SAE J3016 standard \cite{sae2021j3016}, which describe conditional to full driving automation and are the primary focus of both industrial deployment and academic investigation. To address this, we present a large-scale survey and literature review on two systems under test (SUT): modular and end-to-end (E2E) ADSs. Modular ADSs are composed of separate perception, planning, and control modules, while E2E ADSs rely on a unified deep learning model to directly map sensor inputs to driving actions. This study aims to explore three key research questions (RQs): 
\begin{itemize}
    \item \textbf{RQ1:} What are the current practices and demands in ADS testing?
    \item \textbf{RQ2:} To what extent does the current work contribute to the industry practitioners' and academic researchers' demands?
    \item \textbf{RQ3:} What are the challenges and future directions of demands?
\end{itemize}

Our survey methodology has three stages: discussion, survey, and follow-up. Specifically, we first discuss current practices and demands with professionals to acquire an initial understanding. Based on these, we design a 107-question survey and distribute it to gather industry practitioners' and academic researchers' experiences and expectations. Finally, in the follow-up stage, participants explain the reasons for their demands, helping us explore current demands in depth.

Our survey studies three aspects: the basic information about ADS and its testing techniques, testing on V2X communication, and the utilization of FMs in ADS testing. Based on the response, we summarized 7 demands, including 1 V2X-related and 2 FM-related ADS testing demands. We analyzed the perspectives of both practitioners and researchers on each demand, highlighting their respective concerns and priorities. The demands are (1) diversity in corner cases, (2) bridging gaps between simulation and real-world testing, (3) more comprehensive testing criteria, (4) effective defenses against current attacks, and (5) seamless cross-model collaboration in V2X systems. The demands of FMs for testing contain (6) Test case quality generated by LLMs and (7) reliable cross-modality integration when using FMs. To the best of our knowledge, we are the first to conduct surveys on the V2X system and utilization of FMs in ADS testing. To understand the state-of-the-art techniques in ADS testing, we conduct literature reviews on papers published in mainstream venues across various domains, including software engineering (TSE, TOSEM, ICSE, ISSTA, FSE, ASE, etc.), vehicle intelligence (TITS, TIV, IV, etc.) and artificial intelligence (AAAI, CVPR, NeurIPS, etc.) Along with the survey responses, we conclude seven emerging challenges and demands. 

\textbf{Contribution.} The main contributions of this work are as follows:
\begin{itemize}
    \item We conduct an advanced ADS testing survey by investigating testing practices and challenges in two SUTs: modular and E2E ADS.
    \item We are the first to extend the ADS testing survey by investigating V2X collaboration and the role of FMs in ADS testing, discussing their challenges and the gaps in current testing frameworks.
    \item We conduct a large-scale survey (107 questions) with both academic and industrial participants, identifying key testing demands and gaps between research and real-world practices.
\end{itemize}

\section{Related Work} \label{section2}
Huang \textit{et~al.} conducted a comprehensive literature review of the safety of Deep Neural Networks (DNNs) by surveying 202 papers \cite{huang2020survey}. They discussed the testing of DNNs and identified 11 prominent challenges. Wang \textit{et~al.} surveyed software testing with LLMs by analyzing 102 relevant papers and summarizing several challenges, such as the case coverage and real-world application \cite{wang2024software}. Compared to their work, our study focuses on ADS testing by not only conducting literature reviews but also a large-scale survey. 

Many literature reviews centre around ADS testing reviews \cite{song2024empirically, song2024industry, tang2023survey, lou2022testing}, including scenario-based ADS testing \cite{song2024empirically}, critical scenario identification and challenges \cite{song2024industry}. Compared to these reviews, we conducted a comprehensive survey of the ADS testing and discussed the general testing challenges. Among those overall reviews, the work from Tang \textit{et~al.} presented a thorough literature review of ADS testing and identified the challenges and opportunities \cite{tang2023survey}. Compared to their work, our work invited 100 ADS testing practitioners or researchers to participate in our survey and follow-up open-ended questions to analyze the current practices and challenges of ADS testing. The reviews conducted by \cite{lou2022testing} also focus on ADS testing and are most related to our study. However, they do not discuss the V2X system and the utilization of FMs in ADS testing, which we include in our paper.

\begin{figure*}
    \centering
    \includegraphics[width=1.0\linewidth]{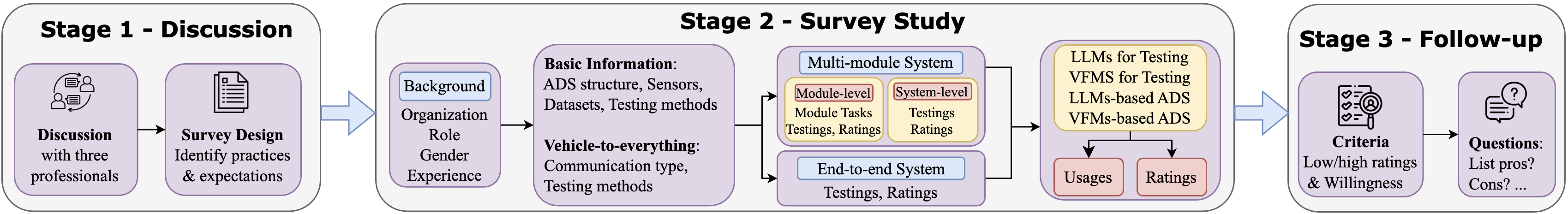}
    \caption{The process of our survey.}
    \label{fig:methodology}
\end{figure*}

\section{Methodology} \label{section3}
Following practices in \cite{shull2008guide, lou2022testing}, our survey design contains three stages: \textbf{Discussion}, \textbf{Survey} and \textbf{Follow-up} as displayed in Figure \ref{fig:methodology}. We first discussed with four professionals in the ADS industry to initially understand the current practices and expectations for ADS testing. Then, we designed a comprehensive and large-scale survey of 100 ADS testing practitioners and researchers. For participants who meet our follow-up criteria, we further invited them to answer open-ended questions.

\begin{table}[]
    \normalsize
    \centering
    \caption{Professionals on survey design discussion stage.}
    \resizebox{0.48\textwidth}{!}{
    \fontfamily{ptm}\selectfont
    \begin{tabular}{cccc}
    \toprule
    Professionals & Organization & Role & Experience \\
    \hline
    P1 & Company A & ADS Testing Engineer & 5 years \\
    P2 & Company B & ADS Testing Engineer & 3 years \\
    P3 & Research Institute A & ADS Software Engineer & 4 years \\
    P4 & Research Institute A & ADS Testing Engineer & 3 years \\
    \bottomrule
    \end{tabular}}
    \label{tab:practitioners}
\end{table}

\subsection{Discussion} 
We invited four professionals from two automobile companies and one research institute specializing in ADS to discuss current practices, challenges, and future opportunities in ADS testing. They bring extensive experience from collaborations with top-tier automobile manufacturers. Their roles and expertise are outlined in Table \ref{tab:practitioners}. Three are experienced ADS testing specialists, with one having five years and two with three years of involvement in ADS research and development. The fourth is a senior software engineer with four years of experience designing and optimizing ADSs.

The discussions took place across three sessions totaling six hours, where professionals provided in-depth analysis and perspectives on ADS testing. In the first session, we introduced our research motivation and then discussed current ADS testing practices, including sensors, datasets, V2X experiences, classical methods, relevant definitions, and their feedback on testing methods. In the second session, we explored tasks of modular and E2E ADSs, testing approaches, and the implementation of FMs \cite{bommasani2021opportunities} (i.e., LLMs and VFMs) in two aspects: their use in testing techniques and integration into ADSs. Afterward, we reviewed the literature to cover topics not addressed in the first two sessions, such as potential attacks and defenses. The third session focused on these missing aspects and summarized the future direction of ADS testing.

\subsection{Survey} 

\textbf{Survey design.} Based on the discussion, we designed a survey with 107 questions divided into six sections: \textit{Background}, \textit{Basic Information}, \textit{V2X Practices}, \textit{Testing Practices}, \textit{FMs Utilization}, and \textit{Follow-up} (Figure \ref{fig:methodology}). The background section covered participants' organizations, roles, genders, and work experience in ADS testing. The second section focused on the ADS they work with, including sensors and datasets. The third section gathered V2X experience. In the fourth section, we introduced the multi-module and E2E ADS structures. Participants were asked to select the ADS they were most familiar with based on their experience. Those familiar with both structures were asked to choose one to continue the survey. After selecting the multi-module structure, participants were asked about module-level or system-level testing, with deeper questions on tasks and methods for testing.

Specifically, we asked participants to present their testing techniques, including black-box (model inputs and outputs) \cite{sun2020towards}, white-box (full access) \cite{pei2017deepxplore}, and grey-box (partial access) \cite{moukahal2021boosting}. Based on the discussion and literature \cite{song2024industry}, common methods for generating testing scenarios include knowledge-based \cite{deng2022declarative}, search-based \cite{huai2023sceno}, and data-driven \cite{zohdinasab2024focused}. Knowledge-based methods use expert knowledge, search-based methods employ algorithms (e.g., genetic algorithm \cite{kluck2023empirical}), and data-driven methods rely on real-world data. A more detailed literature review is provided in Section \ref{literature}. Participants rated the methods on a 7-point Likert scale \cite{joshi2015likert}, from "Strongly disagree" (level-1) to "Strongly agree" (level-7). Those familiar with both module-level and system-level testing answered questions for both. Respondents with E2E experience were asked about sensors, output control actions, and testing ratings. The fifth section explored their experience with FMs, including usage and effectiveness compared to traditional methods. In the final section, participants could provide their email to receive study results and were invited for follow-up. "Other" options were included for each multi-choice question to capture diverse responses. Additionally, some survey questions allowed multiple selections to capture the real-world use of combined testing approaches. For example, participants could select more than one testing strategy (e.g., black-box, white-box, gray-box) or scenario generation method (e.g., knowledge-based, search-based, data-driven), reflecting hybrid practices commonly adopted in ADS testing.

\noindent \textbf{Participants.} We invited participants using two approaches: (1) personal networks and (2) emails crawled in GitHub. Specifically, our personal networks included individuals from both academia and industry, such as partner ADS research institutions and commercial companies. This allowed us to ensure a diverse pool of respondents in terms of organizational background and professional role. Ten pilot participants from these networks completed the questionnaire to help us identify and mitigate any sensitive or ambiguous questions.

To reach a broader audience, we further identified 48 keywords relevant to ADS testing and used GitHub's API to extract publicly available email addresses of contributors to related repositories. In total, we sent 5,200 emails and received 206 responses. However, due to the large scale of the survey, many responses were incomplete. Moreover, responses from professionals not involved in testing activities (e.g., vehicle dispatchers) were excluded.

The first two authors independently reviewed all responses and each selected 90 complete and high-quality answers. Among these, 88 responses overlapped, resulting in a 97.78\% selection consistency. The final 90 responses were determined through discussion and consensus. Including the 10 pilot responses, we obtained 100 valid and complete survey responses for analysis. Regarding participant demographics, the majority (59\%) are from research institutes, with research scientists (43\%) being the largest professional group. The male-to-female ratio is approximately 8:2, and 73.3\% have 1 to 5 years of experience in ADS testing. The respondents include both researchers and practitioners, allowing us to compare perspectives across different roles. We explicitly differentiate their responses throughout our analysis to highlight the varying concerns and testing priorities from both communities.

Additionally, participants span eight countries, with the majority from China (21 participants), the USA (17), and Japan (12). These respondents were primarily recruited through our professional network, which includes collaborators from both academic institutions and industry organizations in these countries. Together, they represent over half of our sample and cover key ADS markets with diverse regulatory and deployment paradigms: China and the USA emphasize rapid field application and commercialization, while Japan and several European countries prioritize safety and cautious deployment strategies \cite{alqahtani2025recent}. Although our survey may not fully reflect every global regulatory environment, the inclusion of participants from these major regions ensures that our findings capture broadly relevant practices and challenges in ADS testing.

\noindent \textbf{Data Analysis.} In this study, we counted and summarized all the data collected in the survey in the form of text or graphs. In the survey, we placed ``Other'' options for each multiple-choice question to prevent the provided options from not covering the participants' experience. We also analyzed this part of the alternative answer. If the answer belonged to a given option, it was correctly categorized. Otherwise, it was briefly introduced when summarizing the relevant questions. We analyzed participants' demands in linear-scale ratings by comparing the rating distributions.

\subsection{Follow-up}
We conducted follow-up surveys of participants with the following criteria: (1) rating low/high on the 7-point Likert scale, (2) willing to attend the follow-up stage. There were 27 participants in our follow-up phase, and we developed multiple targeted open-ended questions for each of them according to their responses. For instance, \textit{Please specify why the current testing approaches cannot meet the testing requirements}. Regarding the data analysis in this stage, the first two authors categorized and summarized all open-ended questions collected separately to identify current demands further. The results were then discussed with each other to refine the results.

\noindent \textbf{Data Availability.}
The complete survey questions, response, and supplement materials are publically available at \url{https://github.com/ADSTestingSurvey/ADS-Testing}.

\subsection{Literature Review}
We follow the literature review guidelines proposed by \cite{keele2007guidelines} to ensure a comprehensive and structured exploration of relevant research. Our literature search strategy consists of three main methods: keyword search, citation tracking, and journal browsing. 

For the keyword search, we carefully designed 52 keywords based on a preliminary review of ADS testing research and relevant survey papers (e.g., autonomous driving testing, E2E testing, online testing, CARLA, etc.). The keyword list was iteratively refined to ensure broad coverage of relevant studies. The final set of keywords was combined using Boolean operators such as ``OR'' and ``AND'' to expand coverage, and adapted for use in major databases including IEEE Xplore, ACM Digital Library, SpringerLink, and ScienceDirect. Search results were first automatically filtered by removing duplicates and unrelated domains, followed by manual screening based on titles and abstracts. Where ambiguity remained, the full-text review was conducted. To reduce bias, two authors independently screened each candidate paper, resolving disagreements through discussion. This ensured consistency and minimized selection bias in the final corpus of reviewed literature. 

We then employed citation tracking, which includes both forward and backward tracking. Forward tracking identifies papers that have cited a given study, helping us trace its impact and subsequent advancements, while backward tracking involves reviewing the reference lists of key papers to identify foundational and related works. To ensure the robustness of our selection, we prioritize highly cited papers and those published in premier SE, ADS, and AI venues. Forward tracking is conducted using Google Scholar, while backward tracking is performed manually. To further refine our selection, we conduct journal and conference browsing. In particular, we focused on software engineering journals and conferences (TSE, TOSEM, IST, ICSE, ESEC/FSE, ASE, ISSTA, ICST, etc.), and ADS and AI-related venues (TITS, TIV, IV, AAAI, NeurIPS, CVPR, etc.). We prioritized recent papers from the last five years while including seminal works significantly shaping ADS testing research. A detailed discussion of these contributions, existing challenges, and research gaps is presented in Section \ref{literature}.

\section{Common Practices of ADS Testing}  \label{section4}
This section provides an overview of ADS testing practices used by industry and academia. Based on our survey results, academic researchers tend to focus on innovation and novel techniques, while industry practitioners are more concerned with system dependability, robustness, and practical deployment. Reflecting these differences, we examine the testing practices of two types of SUT: multi-module and E2E ADSs, followed by a review of the sensors involved. We then examine testing practices for single-agent ADSs, including sensor testing, and analyze multi-module systems with module-level and system-level testing. We also cover E2E system testing approaches and focus on V2X communication testing for multi-agent and E2E systems. Finally, we highlight the use of FMs in ADS testing and their impact on system performance.

\begin{figure*}
    \centering
    \subfigure[Multi-module system.]{
        \includegraphics[width=0.45\linewidth]{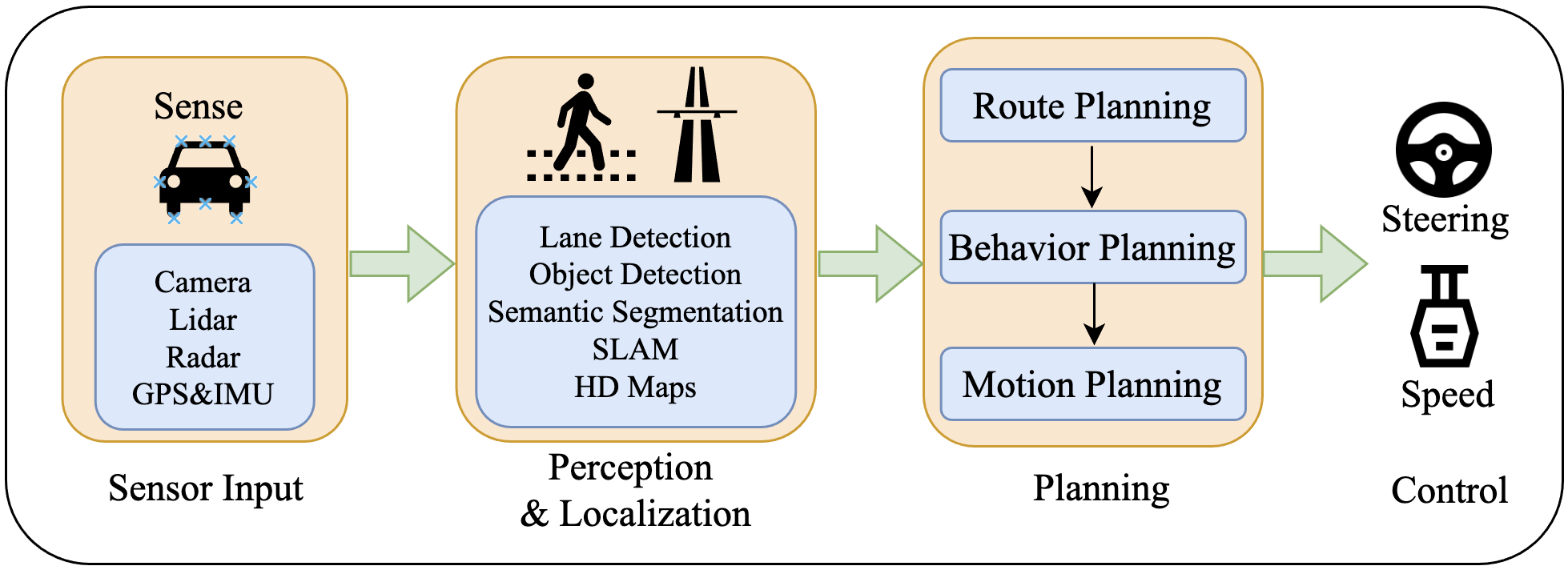}
        \label{fig:multi}}
    \subfigure[End-to-end system.]{
        \includegraphics[width=0.45\linewidth]{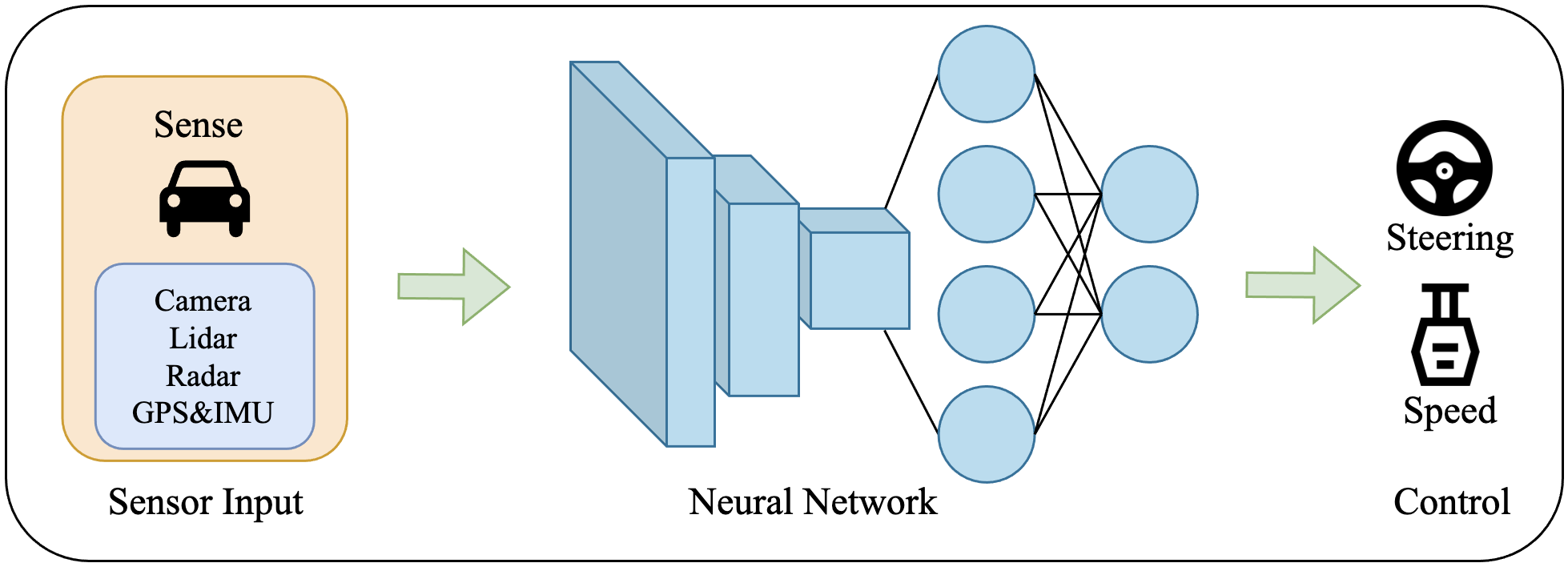}
        \label{fig:e2e}}
    \caption{The structure of the multi-module system and end-to-end system.}
    \label{fig:multi&e2e}
\end{figure*}

\subsection{Autonomous Driving Systems under Test}
Participants in our survey work on two types of ADSs: multi-module and E2E. In this section, we first introduce the definition and structure of these two ADSs and then analyze the survey results, followed by the utilization rate of sensors. 

\noindent \textbf{Multi-module ADSs:} Multi-module ADSs use a modular architecture, decomposing the system into interconnected components, each dedicated to a specific task. This structure is shown in Figure \ref{fig:multi}, enables flexibility, scalability, and easier debugging compared to monolithic designs. Typically, a multi-module ADS includes sensing, perception and localization, planning, and control modules. The Sensing module collects raw data from sensors like cameras, LiDAR, Radar, and GPS to understand the vehicle’s surroundings. The Perception and Localization module processes this data to detect road objects (e.g., vehicles, pedestrians) and localize the ego vehicle. The Planning module then generates an optimal driving trajectory considering road constraints, traffic rules, and dynamic obstacles. Finally, the Control module converts the planned trajectory into vehicle control commands (e.g., steering, throttle, braking) to execute the driving behavior.

Regarding the distribution of models used in ADSs, participants were allowed to select multiple techniques used in their systems, as real-world ADS modules often integrate several approaches simultaneously. Although DL, reinforcement learning (RL), and traditional ML are all categorized as subfields of ML, we analyze them separately in our survey. Because each has distinct application patterns in ADS testing, and they can also be used in combination (e.g., deep reinforcement learning). To capture this flexible and often overlapping usage, we designed the question as a multiple-choice item. This allows respondents to report all applicable methods and enables to analysis of the adoption trends and co-occurrence patterns of these techniques.

\begin{figure}
    \centering
    \subfigure[Reported usage of learning techniques in ADS testing (ML, DL, and RL are shown separately to reflect their distinct applications and overlapping combinations in practice).]{
        \includegraphics[width=0.8\linewidth]{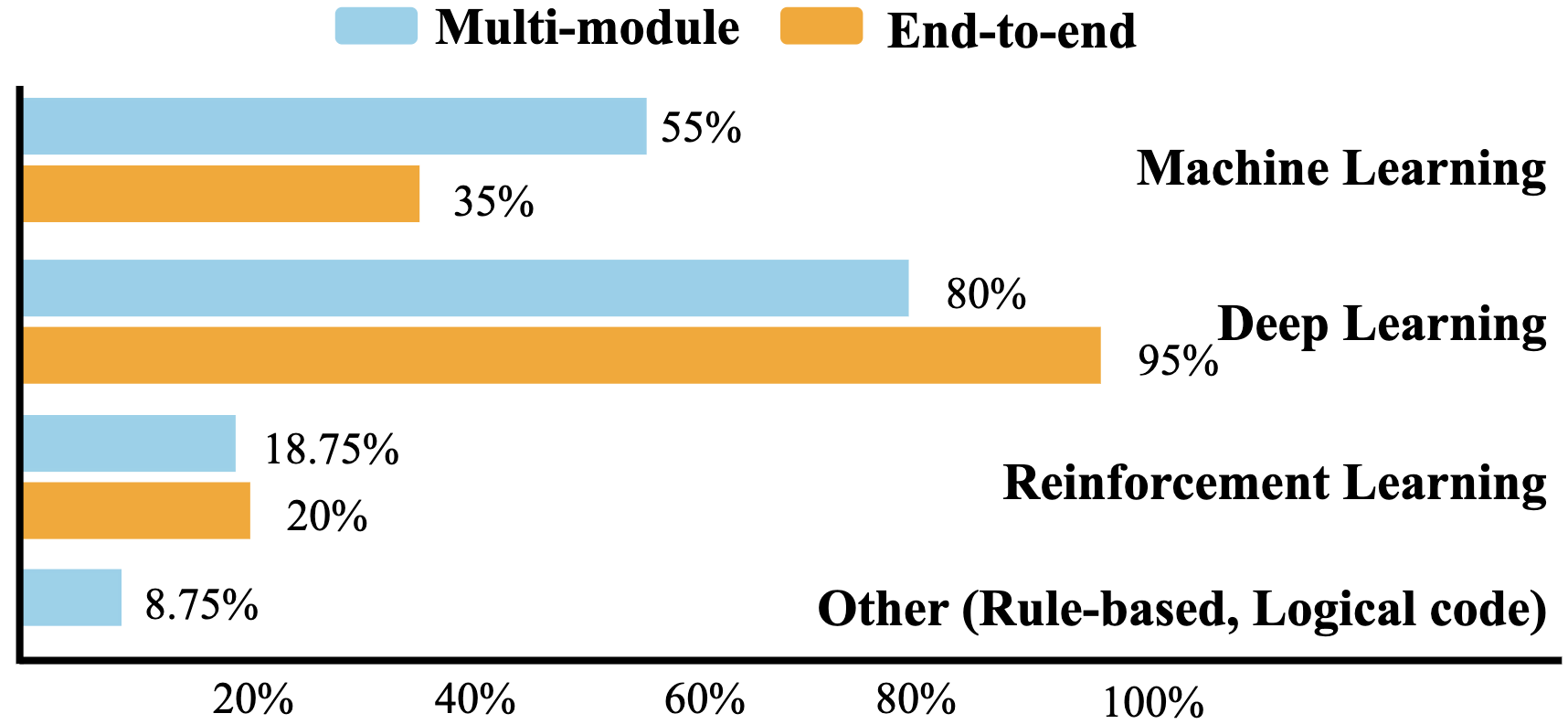}
        \label{fig:archi}}
    \subfigure[Distributions of sensors in ADSs.]{
        \includegraphics[width=0.7\linewidth]{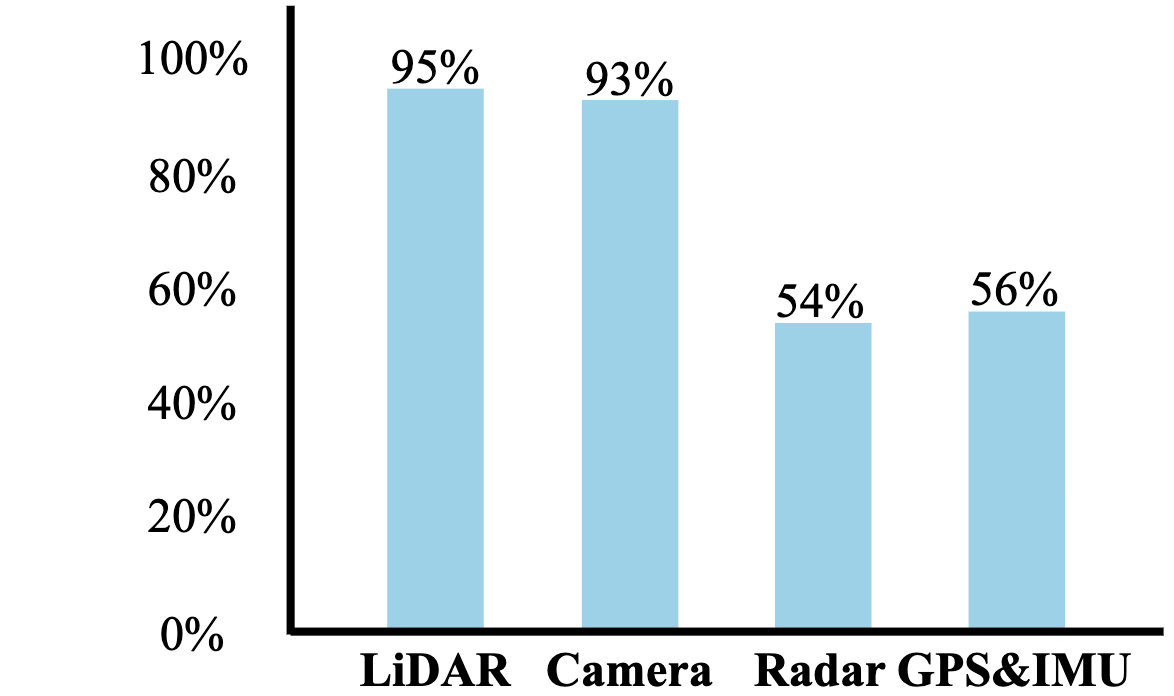}
        \label{fig:sensor}}
    \caption{The survey results related to multi-module and end-to-end systems.}
    \label{fig:ADSs}
\end{figure}

Our survey results show that 80\% participants work on multi-module ADSs, with 45\% from research institutes and 35\% from companies, indicating strong adoption in both academia and industry. As shown in Figure \ref{fig:archi}, 64 respondents integrate DL into their modules (45\% industry practitioners), while 55\% use ML techniques, of which approximately 58\% are academic researchers, including supervised and unsupervised learning models. RL is used by 18.75\% of respondents, half of whom are practitioners. Only 8.75\% use rule-based or logic-based systems, mostly from industry and confined to Planning and Control modules. The rationale behind this trend is discussed in Section \ref{multi-module}, where we analyze the role of hybrid approaches in enhancing ADS performance.

\noindent \textbf{End-to-end ADSs:} The E2E system integrates all modules into a single DNN, mapping sensor data directly to control actions, as shown in Figure \ref{fig:e2e}. Compared to multi-module ADSs, E2E systems have a simpler structure and rely more on data-driven learning, reducing the need for manual rules or feature engineering. Most E2E systems use imitation learning, where DNNs are trained on expert datasets to mimic human driving behavior \cite{lu2023test}. Despite their simplicity, E2E ADSs face challenges in generalization and interpretability, as decision-making is embedded within the network’s representations, complicating debugging and failure analysis. Our survey shows that 20\% of participants work on E2E ADSs, much lower than the 80\% adoption rate of multi-module systems. Among E2E users, 75\% are from research institutes, and 95\% use DNN-based architectures, while ML and RL are only used by academic researchers, with 35\% and 20\% adoption rates, respectively.

Given the heavy reliance on DL models in both multi-module and E2E ADSs, safety concerns, particularly adversarial attacks, have gained attention in autonomous driving \cite{zhang2024uniada}. Among survey participants, 54\% have conducted adversarial testing, with 87.04\% using defenses like adversarial training and model regularization. In addition to adversarial robustness, privacy vulnerabilities in ML-based ADSs are emerging as a concern. One expert noted that ML-based ADSs are susceptible to inference attacks, such as membership and attribute inference \cite{liu2022ml}, which could allow adversaries to infer sensitive user data like locations, restaurants, or driving habits from sensor data.

\begin{tcolorbox}[colback=gray!30, colframe=gray!70]
    \textbf{\textit{Practice 1}}: The majority of participants use DL/ML-based ADSs as system components under test, which are 87.50\% and 45\% on average, respectively, where the adversarial attack is a common threat.
\end{tcolorbox}

All of the professionals expressed their preference for LiDAR and cameras, which can provide precise 3D point clouds and abundant visual information. While radar and GPS\&IMU contribute to specific scenarios, such as collision warning systems, they have limitations in perception and semantic information.

Figure \ref{fig:sensor} shows the distribution of sensor types used in ADSs, based on participants’ multiple-choice responses. LiDAR and cameras are the most widely adopted, with utilization rates of 95\% and 93\%, while Radar and GPS \& IMU are used in approximately 50\% of systems. The sensor utilization is similar among industry practitioners and academic researchers. All industry professionals preferred LiDAR and cameras for their high-resolution depth perception and rich visual information, crucial for object detection, localization, and scene understanding. Radar and GPS \& IMU serve complementary roles, where Radar is valued for its robustness in adverse weather and its ability to measure object velocity, while GPS \& IMU are essential for global localization and vehicle odometry in GPS-denied environments. However, Radar and GPS \& IMU have limitations in perception, resulting in lower utilization rates. Sensor fusion is a key trend, and integrating these sensors effectively remains a research challenge to improve system robustness across diverse driving conditions.

\subsection{Testing for Single-agent ADSs} \label{single-agent}
Next, we focus on single-agent ADS testing. We first summarize the testing distribution of single-agent ADS, followed by presenting the sensor-related findings, and then analyze the testing methods in further detail by dividing multi-module into module-level and system-level testing. Finally, we discuss testing issues related to the E2E system.

\begin{figure}
    \centering
    \subfigure[Distribution of single-agents.]{
        \includegraphics[width=0.45\linewidth]{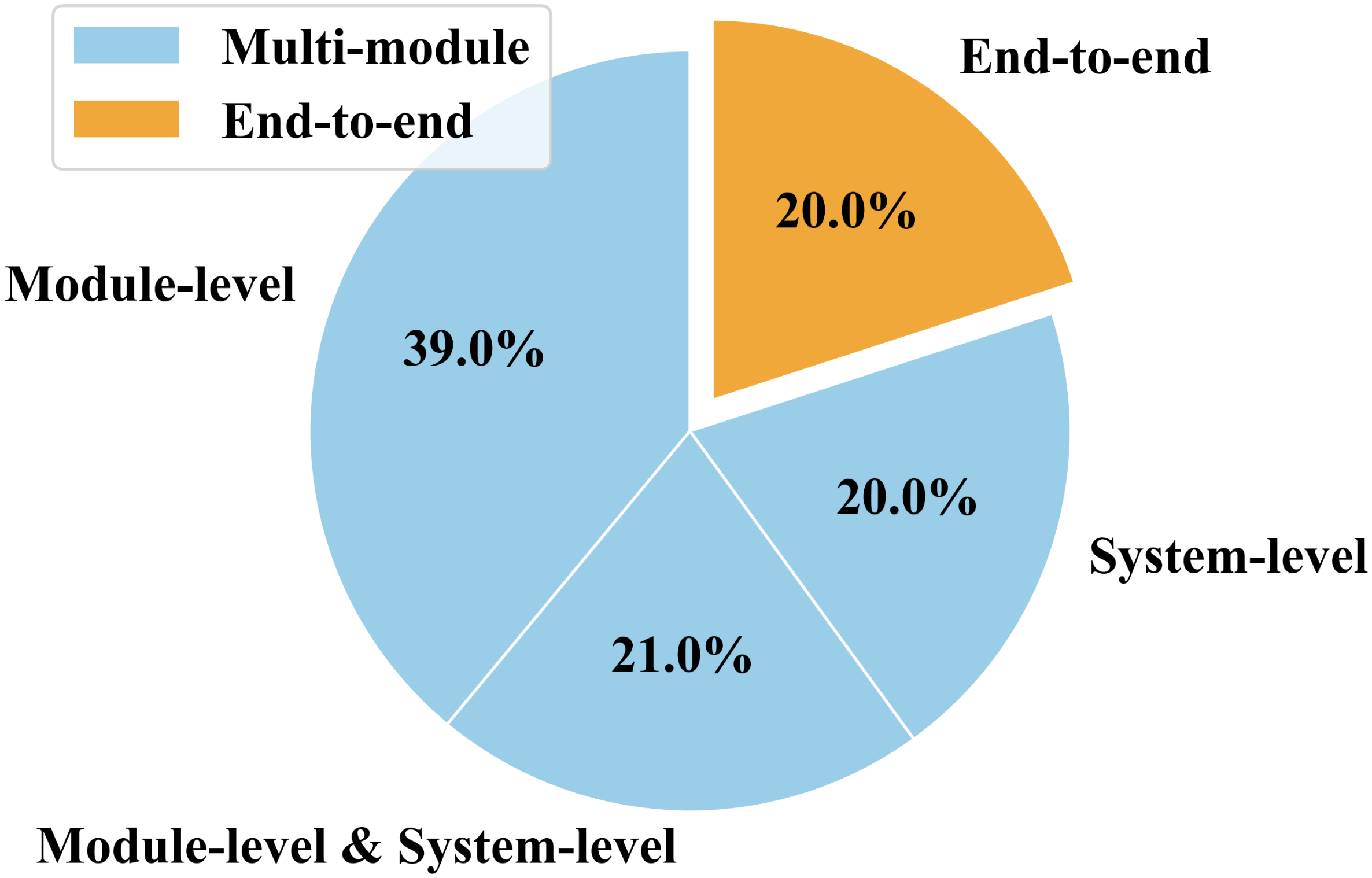}
        \label{fig:N_mul&e}}
    \subfigure[Distribution of simulations.]{
        \includegraphics[width=0.4\linewidth]{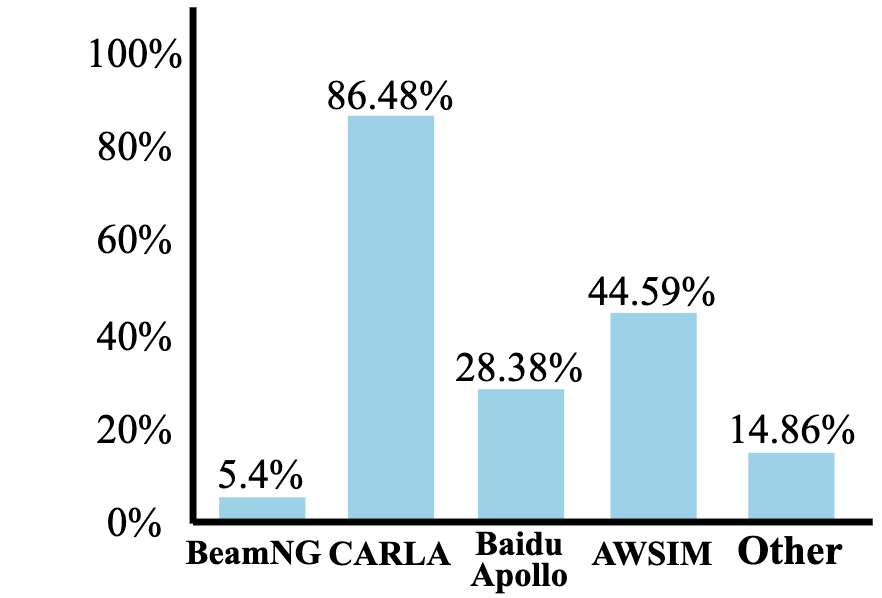}
        \label{fig:simulation}}
    \caption{The survey result of single-agent testing in ADSs.}
    \label{fig:single1}
\end{figure}

Figure \ref{fig:N_mul&e} shows the distribution of testing practices in single-agent ADSs, where multi-module ADSs account for 80\% of participants, with 36\% from industry. 39\% focus on module-level testing (over half from academia), and 21\% conduct both module and system-level testing. We found 52.73\% of industry practitioners focus on real-world testing, while 62.30\% of academic researchers use simulation platforms, likely due to cost and safety constraints in real-world testing. Figure \ref{fig:simulation} shows CARLA \cite{dosovitskiy2017carla} is the most popular simulation platform, praised for its flexibility, high-fidelity urban scenarios, and open-source nature. Since the question allowed multiple selections, 74\% of participants worked on online testing, while 55\% conducted offline testing, with more industry practitioners preferring offline testing and more researchers using online testing for cost reasons. Professional discussions highlighted that system-level testing emphasizes cooperative behavior, requiring online testing for real-time evaluation, while offline testing is used for controlled experiments and reproducible evaluations.

\noindent \textbf{Testing on Sensors.} 27\% of participants reported conducting sensor testing, with physical attacks like jamming and spoofing posing significant risks to ADS perception, leading to sensor failures or misdetections. To mitigate these risks, 85.19\% of respondents, both from industry and academia, utilize multi-sensor fusion and anomaly detection. Multi-sensor fusion enhances robustness by cross-validating data from multiple sensors, while anomaly detection flags or discards corrupted sensor data. Additionally, 30\% of participants use adversarial training, primarily among researchers (59.52\%), to improve model robustness against adversarial sensor data. Beyond algorithms, 18.5\% have implemented physical protection measures, such as shields, to safeguard sensors from interference and tampering. These findings highlight the importance of combining software-based defenses with physical protection to ensure sensor reliability and ADS perception integrity.

\begin{tcolorbox}[colback=gray!30, colframe=gray!70]
    \textbf{\textit{Practice 2}}: The attacks (jamming, spoofing) of sensors have common defenses, including multi-sensor fusion (85.19\%), anomaly detection (85.19\%), adversarial training (30\%), and physical protection (18.5\%).
\end{tcolorbox}

\subsubsection{Multi-module System.} \label{multi-module}
In this section, we present the multi-module ADS, where functional components are separately designed and tested. We discuss the corresponding testing methodologies, including module-level and system-level testing. Module-level testing ensures individual components function correctly in isolation, while system-level testing evaluates the entire ADS's performance, ensuring seamless integration of all modules in real-world scenarios.

\noindent \textbf{Module-level Testing.} There are 75\% participants who conduct module-level testing in the multi-module system, of which 25\% are industry practitioners. Among these participants, 70\% are testing the perception module, 65\% are focusing on the planning module, and 26.67\% working on the control module. 

In our survey, 95.24\% of perception modules utilize DL models for downstream tasks (e.g., object detection), with 9.52\% (mainly industry practitioners) using rule-based, scan-matching, or classical algorithms. Among the respondents, 76.19\% focus on object detection and tracking, while semantic segmentation accounts for around 20\%, highlighting object detection as the foundation of ADS perception. In the planning module, ML-based and DL-based approaches each make up about 40\%, with RL-based approaches at 12.82\%. The ``Other'' option includes rule-based, and the state machine also occupied 12.82\%, indicating that hybrid approaches combining classical and learning-based methods are still prevalent. The most common tasks are path planning (84.62\%) and behavior planning (58.94\%). In the control module, the majority of participants (62.5\%) shared the classical algorithms they used from both academia and industry, such as proportional integral derivative (PID), model predictive control (MPC), or private algorithms. The specific practices we investigated found that all of the control module respondents engaged in steering control, in which 87.5\% and 56.25\% of them also worked on speed and stability control.

\begin{figure}
    \centering
    \subfigure[Distribution of testing approaches in module-level testing.]{
        \includegraphics[width=0.8\linewidth]{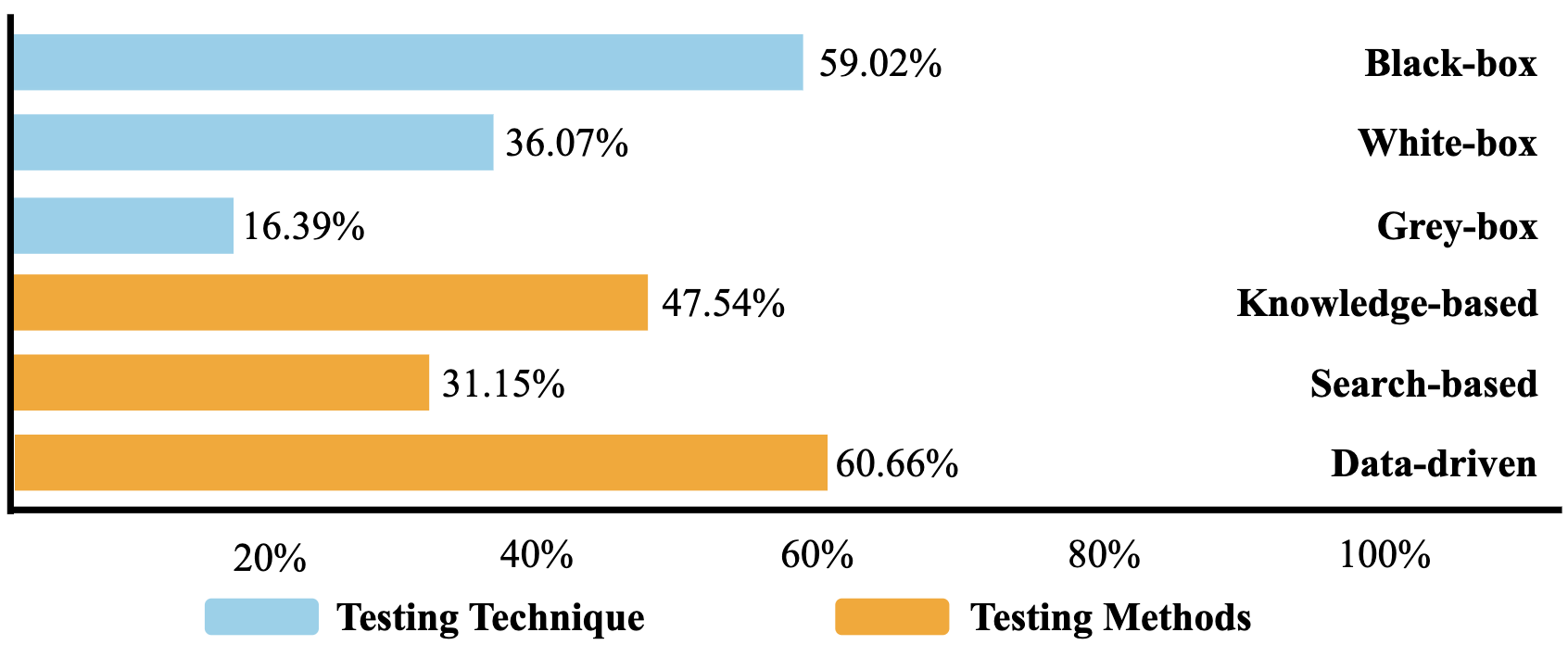}
        \label{fig:multi_method}}
    \hspace{10mm}
    \subfigure[Distribution of testing approaches in system-level testing.]{
        \includegraphics[width=0.8\linewidth]{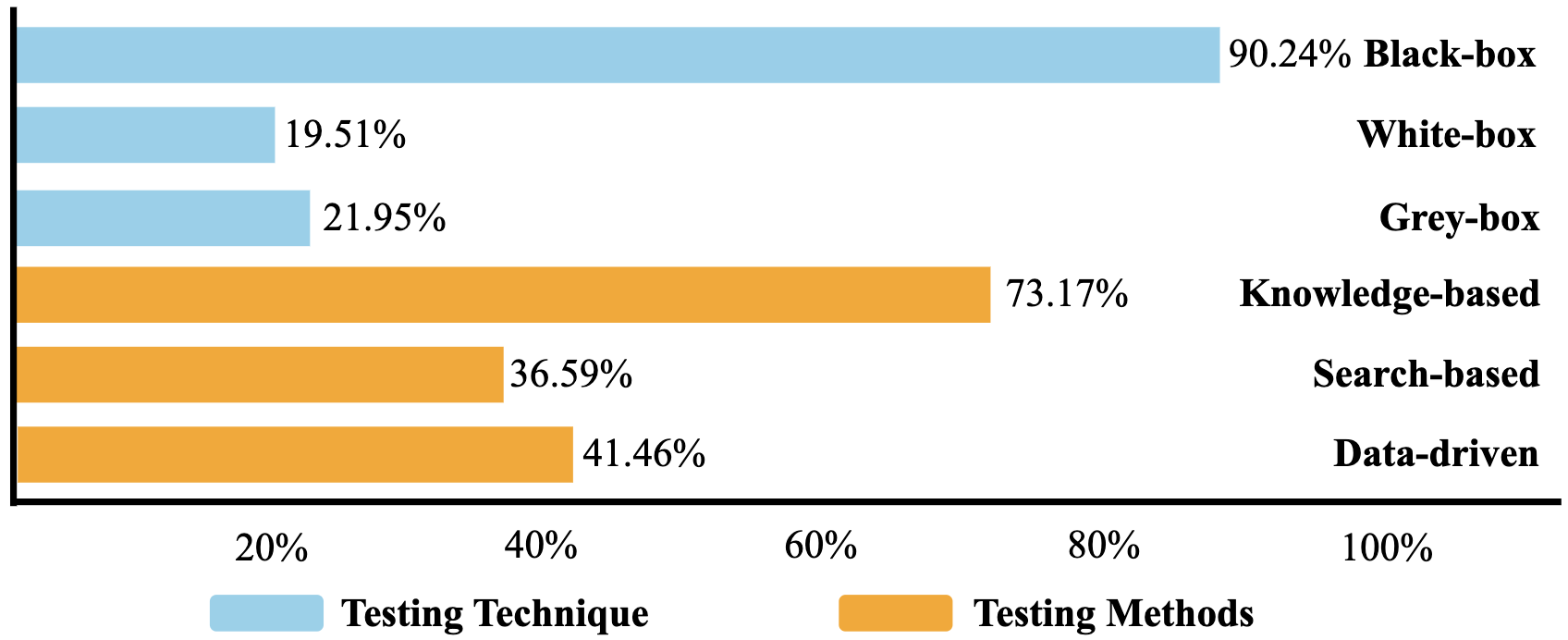}
        \label{fig:sys_method}}
    \caption{The survey result of testing approaches in single-agent systems.}
    \label{fig:system_methos}
\end{figure}

The distribution of testing approaches in module-level testing and system-level testing is shown in Figure \ref{fig:system_methos}, both of which are multiple-choice questions for reflecting real-world testing practices, where different techniques are often used in combination. Not only reveals the popularity of individual methods but also captures how they are jointly applied in practice, offering a more delicate view of ADS testing strategies.

According to Figure \ref{fig:multi_method}, black-box testing is more commonly applied at the module-level testing (60\%), likely due to its ability to evaluate functional correctness and system behavior without requiring internal model access, making it suitable for testing individual components in isolation. Data-driven testing is also prevalent, with 60.66\% of participants adopting this approach, indicating a preference for testing in real-world scenarios or realistic simulation environments to evaluate module performance under diverse and dynamic conditions. Among the three testing approaches, knowledge-based and data-driven methods receive equal attention from both industry and academia, and search-based testing is more prevalent in research institutes.

\noindent \textbf{System-level Testing.}
In system-level testing, black-box testing is the most commonly used method, with a usage rate of 90.24\%, as shown in Figure \ref{fig:sys_method}. This method evaluates the overall performance of ADSs without focusing on the internal details of individual modules. The knowledge-based approach is also widely used, with a rate of 71.79\%, of which 60\% are from industry. Industry professionals emphasize that system-level testing involves evaluating the interaction and integration of multiple modules, which introduces complexity. The knowledge-based method is particularly useful for designing critical scenarios that reflect real-world challenges, such as unexpected road conditions, multi-agent interactions, and rare corner cases. It leverages domain knowledge and predefined rules to explore inter-module dependencies, ensuring safe and reliable operation in diverse environments.

\begin{tcolorbox}[colback=gray!30, colframe=gray!70]
    \textbf{\textit{Practice 3}}: The module-level testing is more flexible in multi-module systems, and both module-level and system-level testing tends to black-box testing, which is 74.63\% on average.
\end{tcolorbox}

\subsubsection{End-to-End System.}
In this section, we present the findings of the E2E system. We introduce the E2E system first, summarizing the output actions utilized by respondents and the testing methodology. The questions in this section also allow participants to select multiple answers. 

The E2E system is an architecture that maps sensor data directly to outputs (e.g., steering, speed, trajectory, etc.). One professional specializing in E2E systems noted that although E2E is less widespread in ADSs, it reduces system complexity compared to multi-module architectures due to fewer interface and integration points. However, we observed diverging views based on participants’ roles. Researchers and algorithm developers often view E2E as simpler, given its streamlined pipeline and lack of inter-module dependencies. In contrast, deployment-focused practitioners emphasized that E2E systems may be more complex in practice. Specifically, the opacity and non-interpretability of DNNs make validation, debugging, and compliance verification more challenging. This contrast highlights how professional backgrounds shape perceptions of architectural trade-offs in ADSs.

\begin{figure}
    \centering
    \subfigure[Distribution of output actions.]{
        \includegraphics[width=0.6\linewidth]{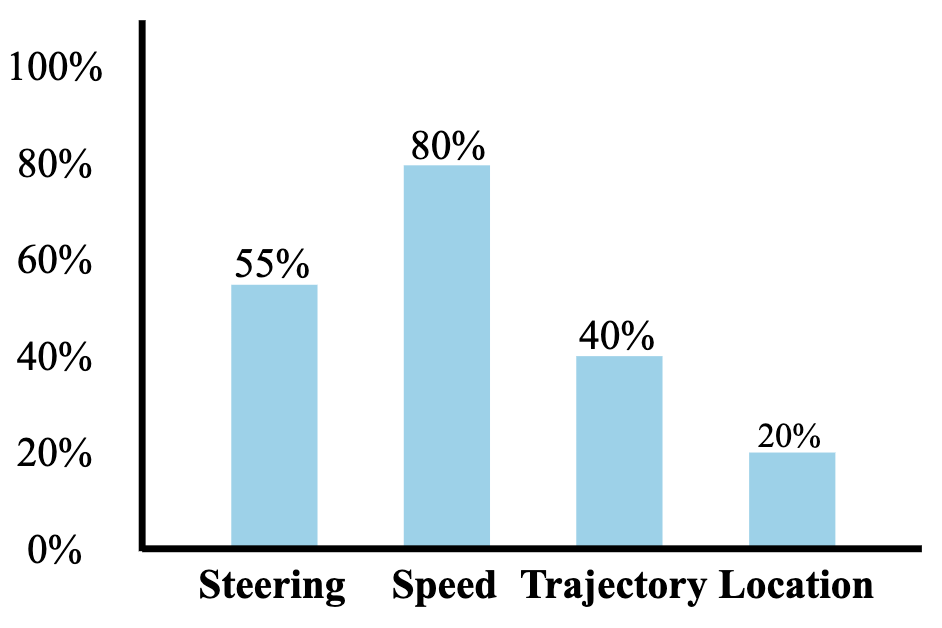}
        \label{fig:e2e_sensor}}
    \hspace{16mm}
    \subfigure[Distribution of testing approaches.]{
        \includegraphics[width=0.8\linewidth]{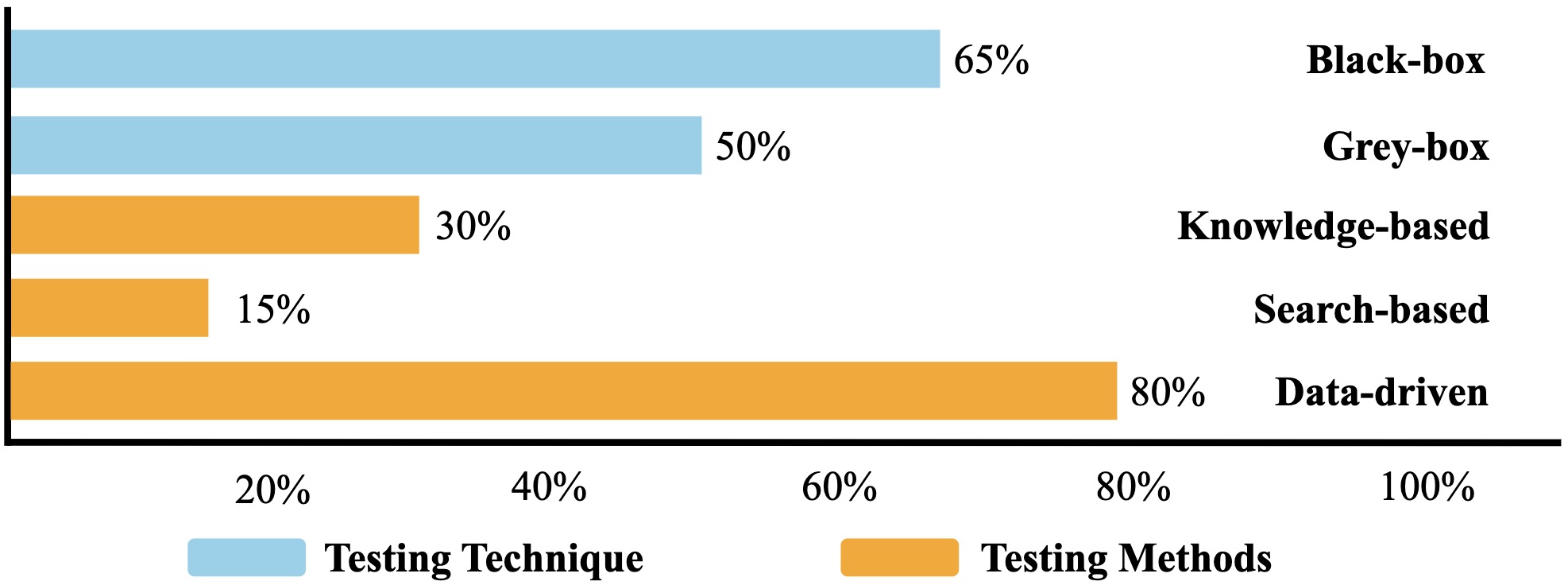}
        \label{fig:e2e_approach}}
    \caption{The survey result of end-to-end system in ADSs.}
    \label{fig:e2e_method}
\end{figure}

Our survey results show that E2E testing mainly focuses on key sensors and output actions, as seen in Figure \ref{fig:e2e_sensor}. LiDAR (70\%) and cameras (70\%) are the most commonly tested sensors, while Radar (30\%) is used less due to its lower spatial resolution. On the output side, speed control (80\%) is the most frequently tested function, followed by steering (55\%), trajectory generation (40\%), and location tracking (20\%), which reflect core vehicle motion control tasks. 

Regarding testing methodologies, our findings (Figure \ref{fig:e2e_approach}) show that E2E testers employ both black-box (65\%) and gray-box (50\%) testing approaches. Black-box testing is useful for evaluating overall system behavior without internal access. While some participants reported applying gray-box testing, this may reflect efforts to improve transparency through techniques such as model interpretability tools, intermediate signal analysis, or limited observability into DNN structures. For example, some practitioners may use feature attribution methods like SHAP to understand model sensitivity to specific input features \cite{kuznietsov2024explainable, nazat2024evaluating}. Others might insert diagnostic checkpoints at intermediate layers of the DNN (e.g., activations from the perception module or intermediate driving intention vectors) to analyze whether internal representations behave as expected during simulation. However, the scope and depth of such testing are still limited compared to conventional gray-box testing in traditional software systems.

Notably, no participant reported using white-box testing, due to E2E systems involving integrated components where the system’s internal workings (such as DNNs and fused sensor data) are not directly accessible. We acknowledge that directly applying white-box or even gray-box testing techniques to neural networks remains an open challenge in the field, and interpretability research may play a crucial role in enabling such testing in the future.

Additionally, the focus is often on testing the system's external behaviors and interactions rather than its internal workings. Among testing techniques, data-driven testing is the most widely used, as illustrated in Figure \ref{fig:e2e_method}. Respondents emphasize the importance of simulating real-world environments and leveraging historical driving data to generate diverse testing scenarios that closely resemble actual deployment conditions.

\subsection{Testing for V2X Communication} \label{approaches}
Following the discussion on single-agent ADS testing, we now turn to the testing of V2X communication systems. V2X communication is essential for enabling effective interactions between ADSs and other vehicles or infrastructure. In this section, we explore how V2X testing varies between multi-agent systems and E2E systems. We begin by defining V2X and then present the survey results regarding the use and testing of V2X approaches across different ADS types.

V2X is a wireless communication technology that enables data exchange between vehicles and external entities, such as other vehicles (V2V), infrastructure (V2I), pedestrians (V2P), and networks (V2N), with the primary goal of enhancing traffic safety and optimizing transportation efficiency \cite{xu2023v2v4real}. By sharing real-time information, V2X improves situational awareness, cooperative decision-making, and collision avoidance, making it vital for the development of ADSs. These benefits drive the need to explore its current use and testing in real-world applications.

\begin{figure}
    \centering
    \includegraphics[width=0.6\linewidth]{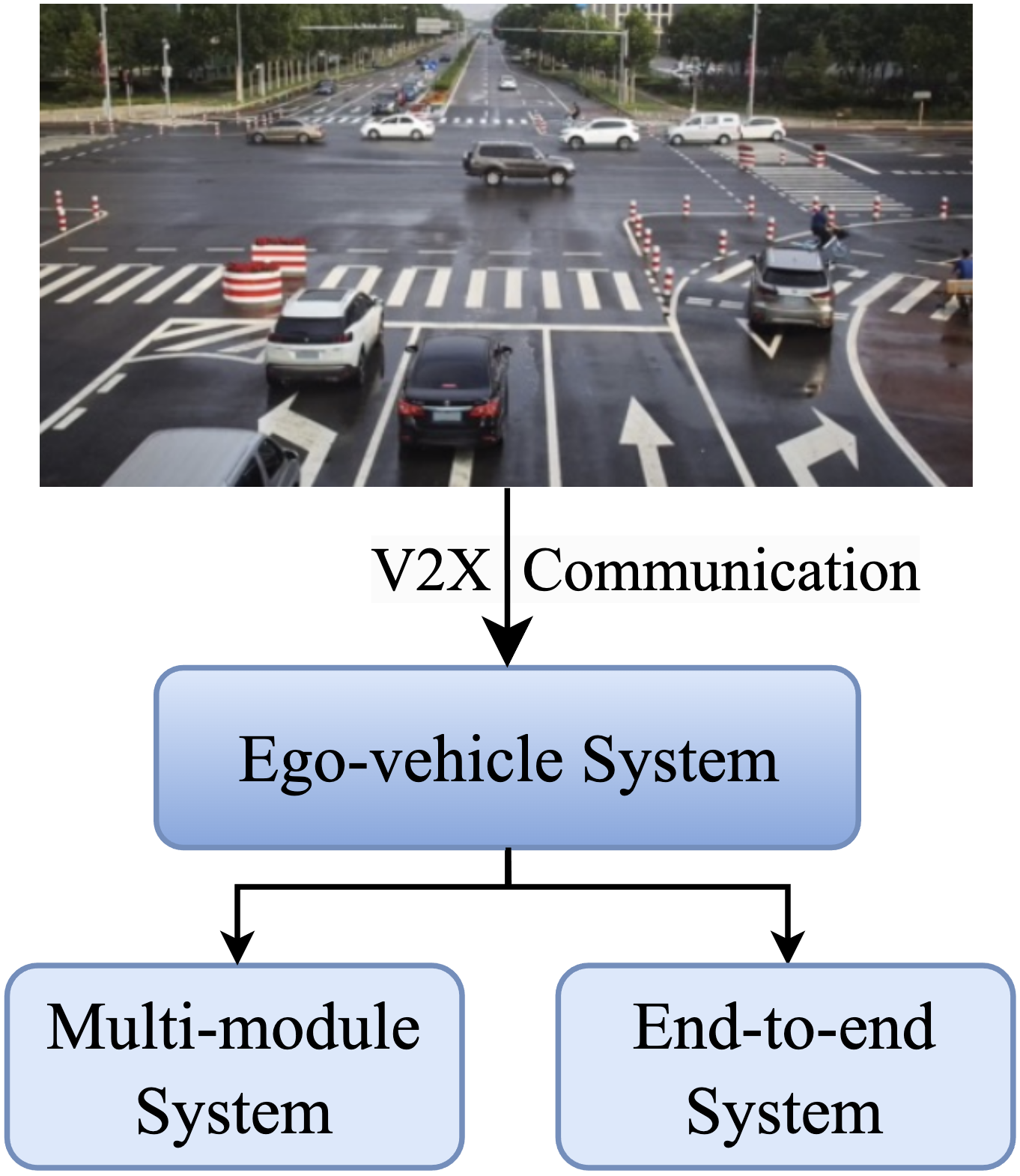}
    \caption{V2X structure.}
    \label{fig:v2x_structure}
\end{figure}

The structure of V2X is shown in Figure \ref{fig:v2x_structure}. A key aspect of V2X-enabled ADSs is data fusion, which integrates information from multiple vehicles and roadside sensors. Data fusion in V2X is categorized into three levels: \textit{early}, \textit{intermediate}, and \textit{late} fusion \cite{xu2022v2x}. \textit{Early} fusion combines raw sensor data (e.g., LiDAR, camera images) from multiple sources before feature extraction, requiring high bandwidth and computation. \textit{Intermediate} fusion uses processed feature representations, while \textit{late} fusion integrates final predictions (e.g., detected objects, planned trajectories), minimizing communication overhead but limiting cross-vehicle feature learning.

The questions in the V2X section are multiple-choice questions. Our survey shows that common V2X datasets include V2X-Sim \cite{li2022v2x} and V2X-Seq \cite{v2x-seq}, which feature communication scenarios between vehicles and infrastructure. Additionally, 36.36\% of V2X users reported conducting cyber security evaluations to assess communication safety and data integrity, mainly in simulated environments. These evaluations focus on threats such as message spoofing, jamming, and data injection. Defense mechanisms include encrypted communication, security protocols, identity authentication, and intrusion detection systems.

\begin{figure}
    \centering
    \subfigure[Distribution of data fusion.]{
        \includegraphics[width=0.75\linewidth]{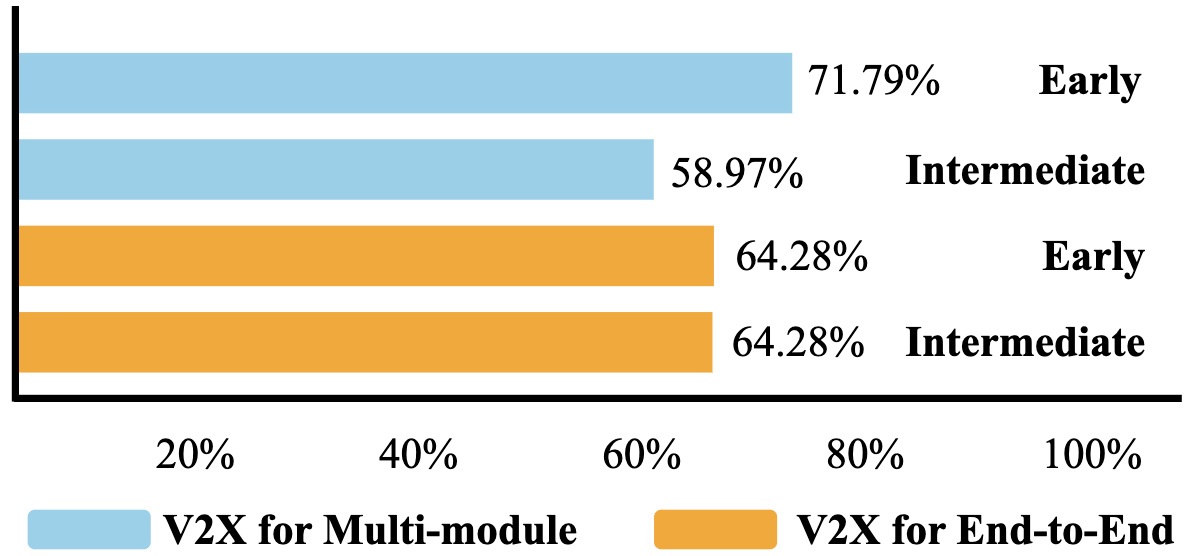}
        \label{fig:fusion}}
    \subfigure[Distribution of V2X testing.]{
        \includegraphics[width=0.8\linewidth]{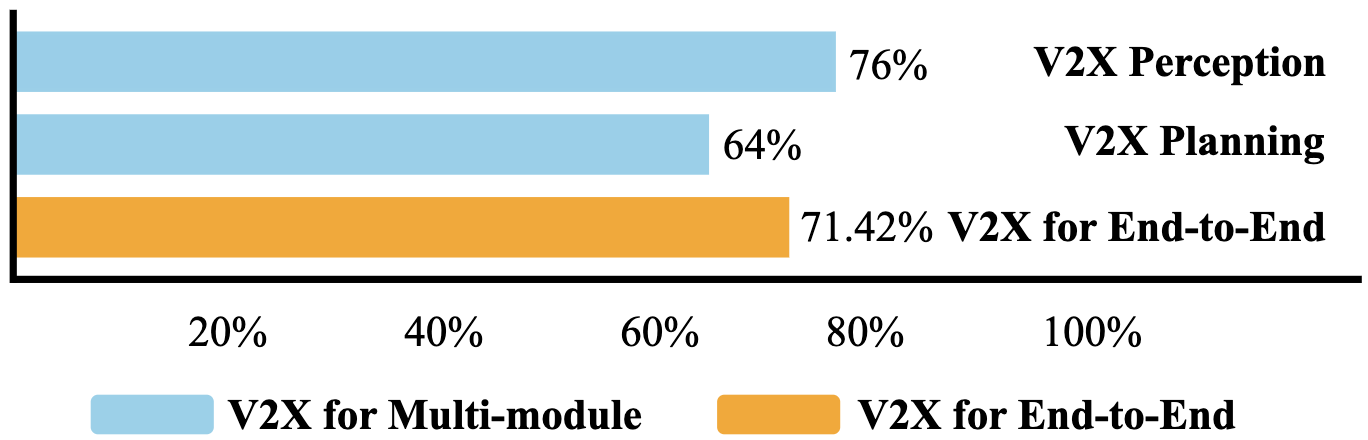}
        \label{fig:N_v2x}}
    \subfigure[Distribution of V2X testing approaches.]{
        \includegraphics[width=0.85\linewidth]{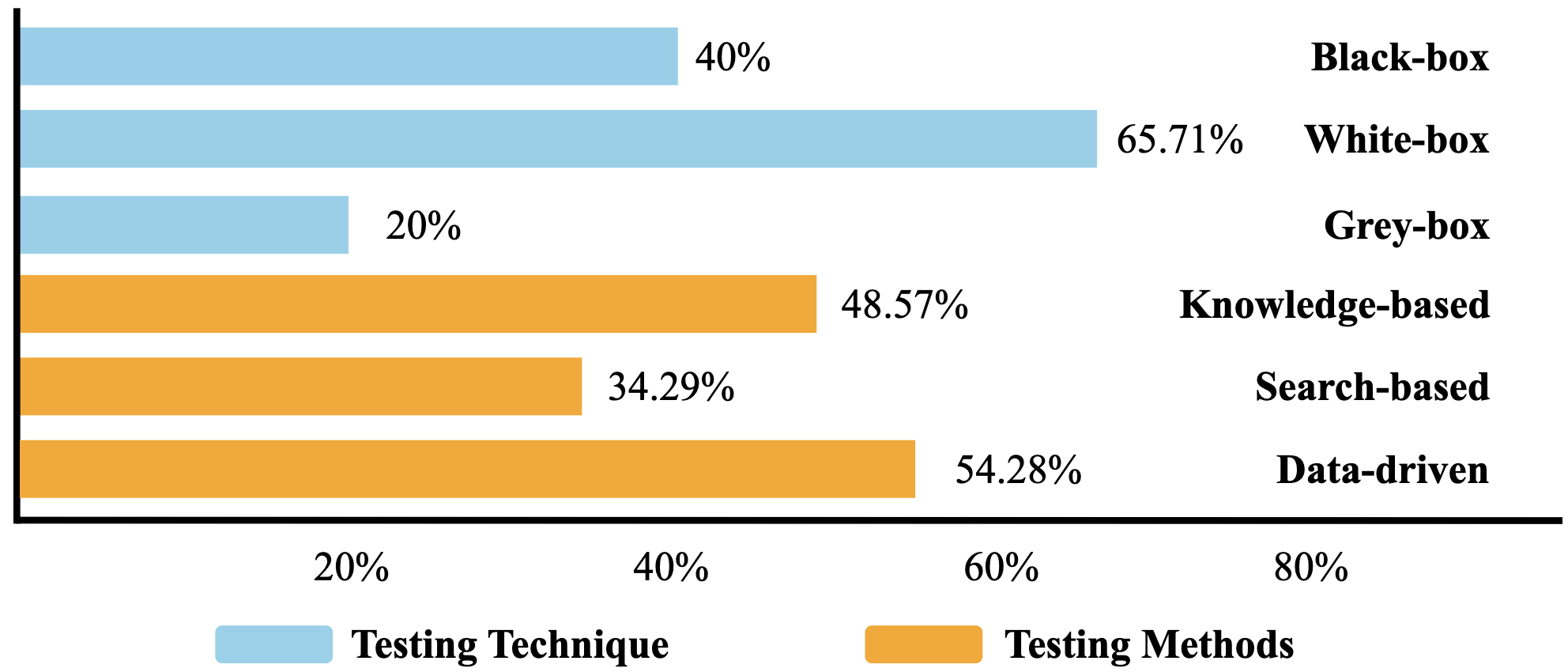}
        \label{fig:v2x_testing}}
    \caption{The survey result related to V2X communication.}
    \label{fig:V2X}
\end{figure}

\noindent \textbf{V2X for Multi-module Agents.} According to our survey, 48.75\% of participants have used V2X in multi-module ADSs, with 60\% from academia, highlighting its growing role in collaborative perception and decision-making. Regarding data fusion, 71.79\% transmitted raw data, 58.97\% used intermediate features, and none used \textit{late} fusion (Figure \ref{fig:fusion}). Specifically, 60\% of those using \textit{early} fusion are from research institutes, while \textit{intermediate} fusion is equally favored by both industry and academia. The preference for early and intermediate fusion reflects the need for rich perception and better scene understanding, while the absence of late fusion suggests a focus on lower-latency, high-fidelity fusion to enhance real-time performance.

From our professional discussion, we identified perception and planning as the primary ADS modules utilizing V2X communication. Among V2X users in multi-module systems, 64.10\% reported conducting V2X-specific testing, with 36\% from industry (Figure \ref{fig:N_v2x}). Specifically, 44\% of academic researchers and 32\% of industry practitioners tested V2X-enhanced perception, while 44\% of researchers and 20\% of practitioners tested V2X-based planning. No participants tested V2X integration in other modules, highlighting its key role in environmental understanding and decision making.

\noindent \textbf{V2X for E2E agents.} In the E2E system, half of the participants stated they used V2X communication, where participants used \textit{early} fusion and \textit{intermediate} fusion are both 64.28\%, which is exhibited in Figure \ref{fig:fusion}. The preference for these fusion strategies aligns with the nature of E2E systems, which rely on rich sensory data and learned feature representations for autonomous decision-making. Regarding testing-related findings, according to Figure \ref{fig:N_v2x}, 71.42\% of V2X users in E2E systems have conducted testing, of which 64.29\% are academic researchers. Our survey shows that 100\% of participants tested LiDAR point cloud transmission, making it the most widely adopted V2X-perceived modality. 80\% utilized RGB images, highlighting the importance of visual perception, while 40\% incorporated radar point clouds, with half from industry, likely due to their robustness in adverse weather conditions.

The testing techniques result is shown in Figure \ref{fig:v2x_testing}. Black-box testing is the most prevalent (40\%) used by all E2E testers for overall system evaluation without internal access. 65.71\% of multi-module V2X testers use white-box testing to analyze module interactions and verify logic compliance. No participant conducts white-box testing in E2E ADSs because it requires deep knowledge of the internal structure of the system, which may not always be feasible or practical in V2X systems due to the complexity of interactions between various vehicle models, sensors, and communication networks. Additionally, 20\% of V2X testers have conducted gray-box testing, the majority of whom are affiliated with research institutes. This type of testing is primarily employed for assessing model interpretability and robustness. In certain cases, gray-box testing also encompasses partial inspections of communication modules or sensor fusion outputs.

In terms of test generation strategies, data-driven (54.28\%) and knowledge-based (48.57\%) are the most frequently employed. Besides, the use of knowledge-based and data-driven approaches is evenly distributed between industry and academia, while search-based methods are predominantly adopted by academic researchers (75\%). On the basis of the professionals' statements, the data-driven method can meet the requirements of V2X systems that need to process large volumes of real-time data, and the knowledge-based method contributes to infrequent corner cases, such as edge cases in cooperative perception, or communication failures.

\begin{tcolorbox}[colback=gray!30, colframe=gray!70]
    \textbf{\textit{Practice 4}}: Most participants leverage early data fusion in V2X communication and conduct testing on perception and planning modules by leveraging the black-box testing technique (65.71\%) and the data-driven method (54.28\%) most frequently. 
\end{tcolorbox}

\subsection{Utilization of FMs}
This section introduces LLMs and VFMs, discusses FMs' utilization for testing, and also slightly introduces survey results on FMs-based ADS, which use VFMs and LLMs for tasks like scenario generation, test case evaluation, and system validation. Survey questions in the FMs sections also allow participants to select multiple choices to share their experiences. 

LLMs are rapidly growing DL-based NLP models with powerful generative capabilities in autonomous driving \cite{cui2024survey}. Traditional ADS testing approaches struggle with the diversity of real-world driving scenarios, but FMs, especially LLMs, can generate diverse test scenarios to improve test coverage and adaptability \cite{zhang2024chatscene, chen2024driving}. Besides, VFMs are models for describing and modeling vehicle characteristics, which are utilized broadly in ADSs \cite{wei2024stronger}. For instance, ADS perception models are highly susceptible to environmental variations (e.g., lighting changes, occlusions, adversarial perturbations), and VFMs can provide a robust way to simulate complex sensor inputs, allowing for stress testing of ADS perception modules under diverse and adversarial conditions. Additionally, they enable data augmentation for synthetic dataset generation, helping improve model generalization and robustness \cite{caron2021emerging}.

\noindent \textbf{FMs for Testing ADSs.} According to the discussion results, all four professionals have relevant experience in testing with FMs, and one has both experience using LLMs-based and VFMs-based ADSs. Professionals stated that LLMs and VFMs both contribute to generating critical scenarios in testing, while VFMs can help retrain by providing real-time feedback from the vehicle. 

\begin{figure}
    \centering
    \subfigure[Distribution of using FMs for testing in multi-module system.]{
        \includegraphics[width=0.7\linewidth]{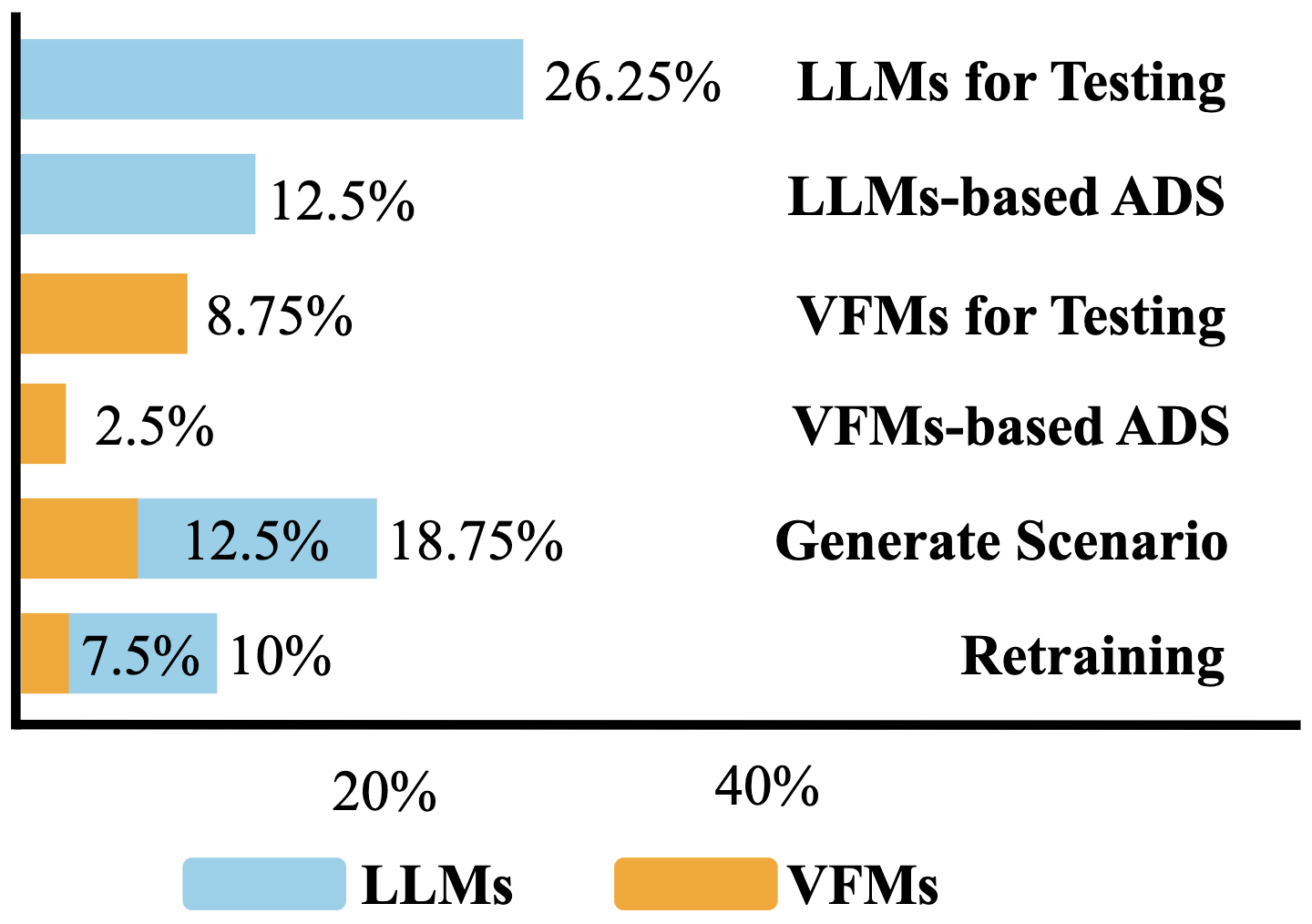}
        \label{fig:LLM_multi}}
    \subfigure[Distribution of using FMs for testing in end-to-end system.]{
        \includegraphics[width=0.7\linewidth]{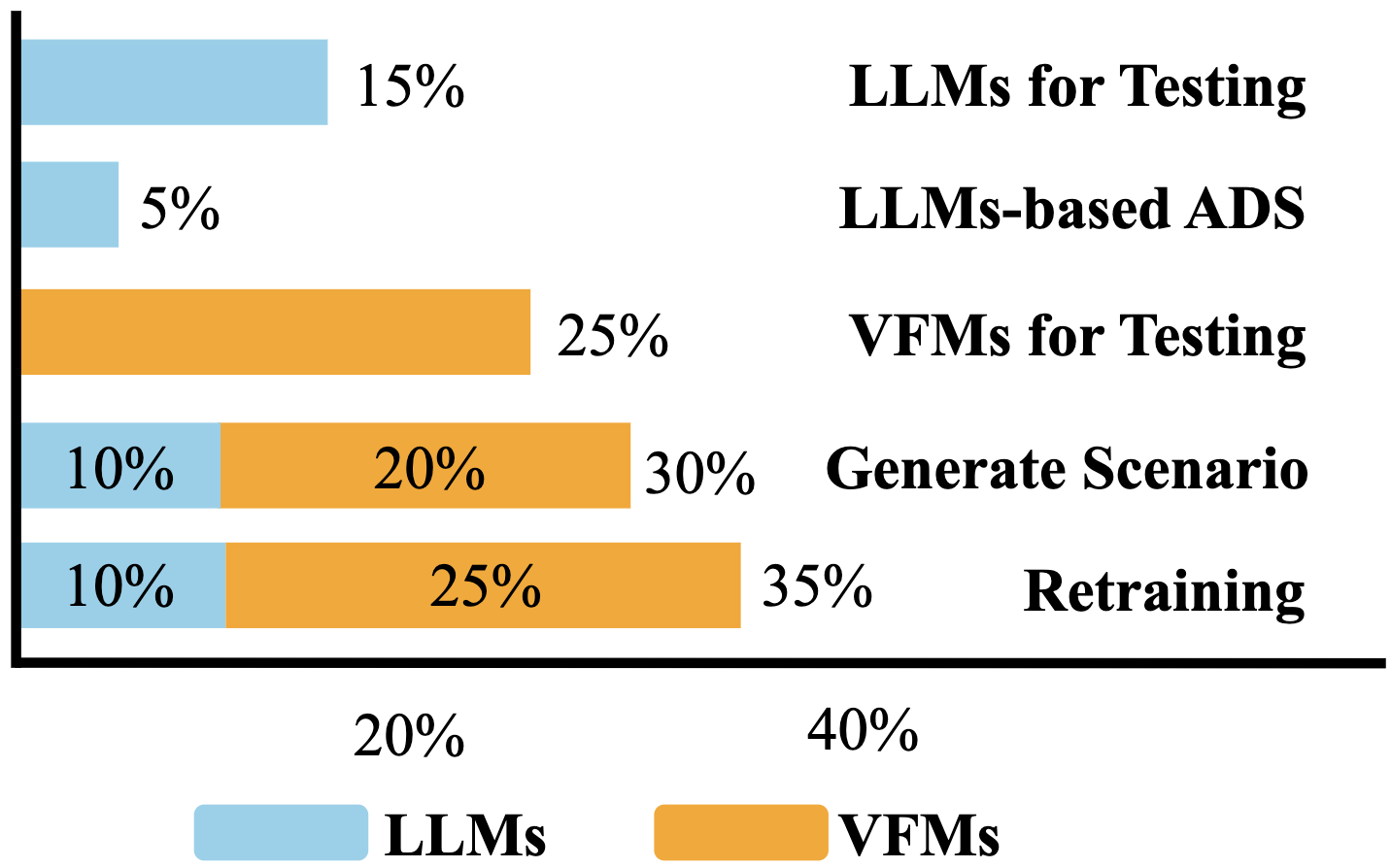}
        \label{fig:LLM_e2e}}
    \caption{The survey result of FMs in ADSs.}
    \label{fig:LLM}
\end{figure}

Survey results show that 70\% of respondents are from research institutes. 26.25\% use LLMs in multi-module ADS testing, while 12.5\% use VFMs, as shown in Figure \ref{fig:LLM_multi}. Among module-level testers, 63.64\% focus on perception, and 45.45\% on planning. The lower adoption of VFMs suggests their strength lies in system optimization rather than module-level testing. VFMs are primarily used for camera-based perception systems, where they model scene variations and generate perturbations. However, radar and LiDAR-based systems, which rely on different principles, may need alternative approaches for testing.

In the E2E system, we found that using VFMs is more frequent (25\%) than LLMs. Unlike multi-module ADSs, where decision-making is modularized, E2E architectures require rapid, continuous feedback for real-time adjustments, making VFM-driven optimization particularly effective in enhancing response time and control stability. 

Additionally, from Figure \ref{fig:LLM_e2e}, participants are more likely to choose LLMs to generate testing scenarios (67.86\%) due to their ability to synthesize complex environments, while VFMs are preferred for retraining (60.71\%). This aligns with the observation that LLMs are well-suited for generating diverse, scenario-based tests, whereas VFMs are more effective for refining real-time feedback loops. The differences in sensor modality and testing target suggest the need for a more tailored approach when selecting FMs for ADS evaluation.

\noindent \textbf{FMs-based ADSs.} LLMs-based ADSs utilize LLMs for tasks like decision-making, scenario understanding, and human-machine interaction, while VFMs-based ADSs improve perception, scene recognition, and environmental understanding. LLMs focus on reasoning and language tasks, while VFMs enhance visual data interpretation. Emerging models like CLIP \cite{radford2021learning}, when combined with Vision Transformers (ViT) and CNNs, enhance tasks like semantic segmentation and object recognition, crucial for ADS perception. Our survey showed 10\% of participants have integrated FMs into ADS workflows, with only 2\% using them as a core component, likely due to computational and adaptation challenges. Respondents also reported using SAM \cite{kirillov2023segment} to improve semantic segmentation in ADS perception.

\begin{tcolorbox}[colback=gray!30, colframe=gray!70]
    \textbf{\textit{Practice 5}}: Participants tend to employ LLMs to generate diverse and complex testing scenarios in multi-module and utilize VFMs for retraining in E2E systems. 
\end{tcolorbox}

\subsection{Hybrid Testing Practices}
Across the ADS testing survey results, participants often employ a combination of testing techniques in ADS testing, which is reflected in the multi-selection design of our survey question. This design choice was intentional, as it mirrors real-world practices where different methods are combined to meet varied testing goals. For instance, combining black-box and white-box testing allows testers to verify both system behavior and internal logic. Similarly, integrating knowledge-based and data-driven methods helps cover domain-specific edge cases and statistically representative scenarios. These hybrid strategies highlight the practical demands of ADS testing, where achieving comprehensive test coverage often requires the complementary strengths of multiple approaches.

Additionally, we observe that even among researchers, testing practices may vary based on their roles. Those focused on developing novel techniques for future ADSs tend to experiment with exploratory or AI-driven testing approaches, while those responsible for evaluating deployed systems prioritize robustness and regulatory compliance. This variation further supports the adoption of hybrid testing practices tailored to different development stages and responsibilities.

\section{Demands}  \label{demands}
\begin{figure}
    \centering
    \subfigure[Ratings for current testing approaches of single-agent ADSs.]{
        \includegraphics[width=0.85\linewidth]{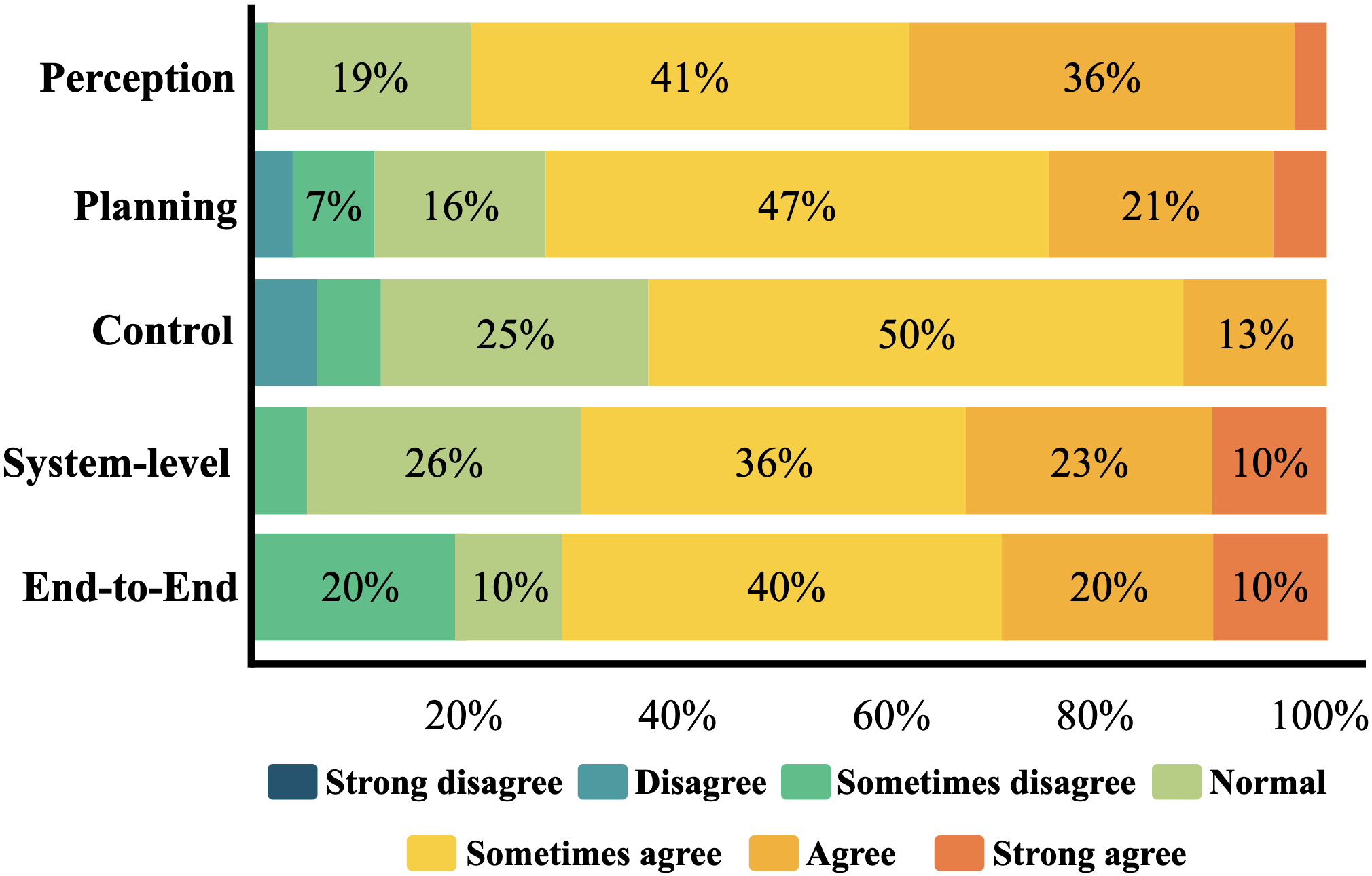}
        \label{fig:rate_common}}
    \hspace{10mm}
    \subfigure[Ratings for LLMs/VFMs testing approaches and LLMs/VFMs-based ADSs.]{
        \includegraphics[width=0.92\linewidth]{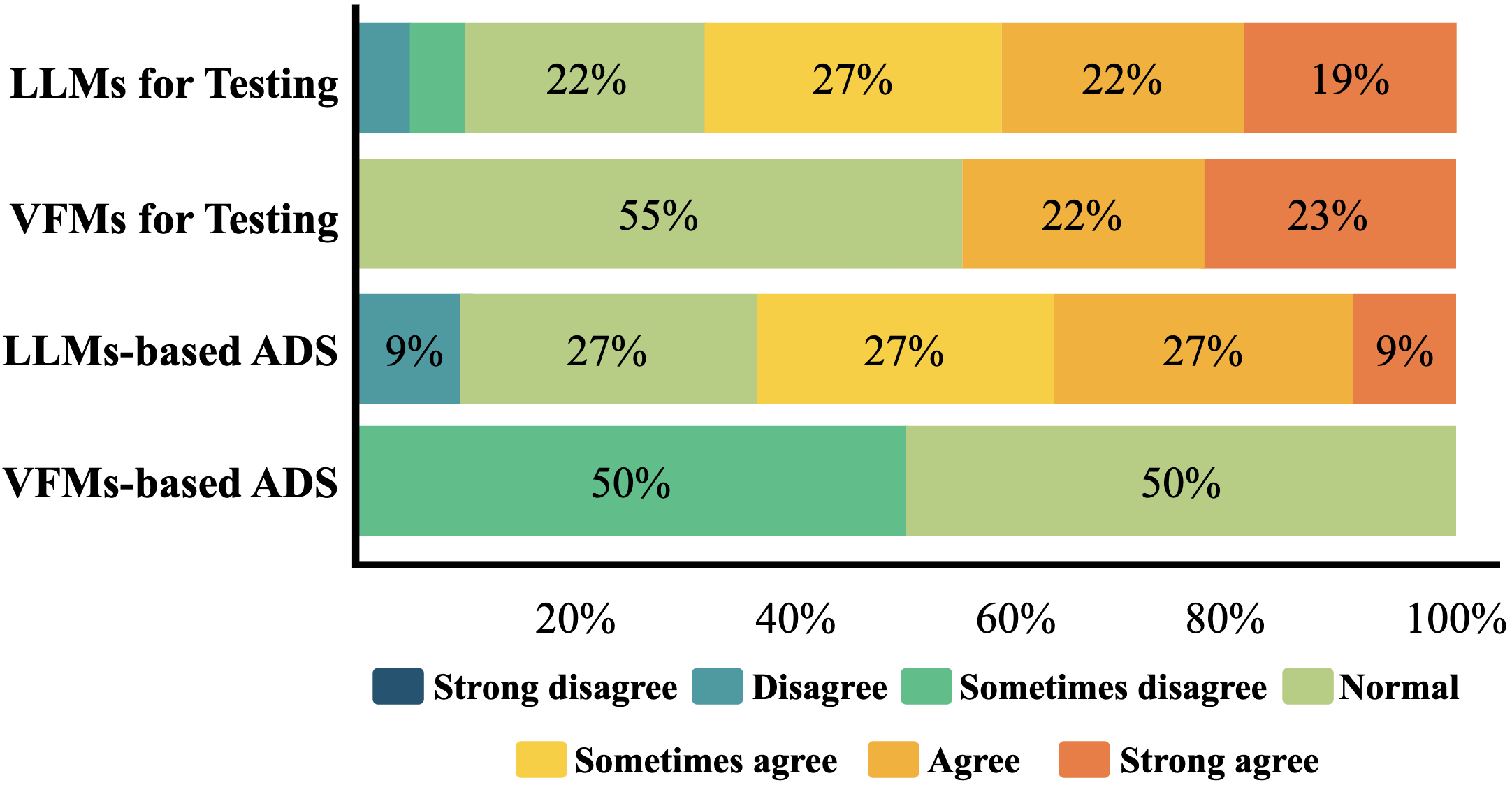}
        \label{fig:rate_LLM}}
    \caption{The rating result of current testing approaches.}
    \label{fig:rate}
\end{figure}

In this section, we summarize 7 demands of ADS testing, including 1 V2X-related and 2 FM-related demands. We analyze demands according to the ratings and the follow-up open-ended questions. The rating result of current testing approaches of single-agent ADSs is illustrated in Figure \ref{fig:rate_common}, which contains a 7-level rate from strongly disagree to strongly agree. According to the result, at the module-level testing, the perception module and the control module received the most and the least positive ratings, 79\% and 63\%, respectively, while 2\% and 12\% of the participants expressed disagreed that the current testing meets the requirements (e.g., safety, robustness, real-time response, user experience), respectively. Besides, no one fully agreed that the current testing of the control module fulfilled the requirements. In the system-level testing, only 5\% of the participants indicated test methods sometimes failed to meet the testing requirements. The E2E system received the most negative voices (20\% sometimes disagree, at level-3).

In summary, 13\% of the participants strongly disagreed or disagreed that their current testing approaches meet their testing requirements. We designed follow-up questions to investigate further (e.g., \textit{Please specify why the current testing approaches cannot meet the testing requirements}), and the response rate reached 84.61\%. 30\% of the participants take a neutral stance on the testing approaches they leverage, the follow-up questions including \textit{Please briefly list the advantages and disadvantages of the current testing approaches}, and received an 80\% response rate. 64\% of the participants sometimes agree on the effectiveness of current testing approaches. The open-ended questions contain \textit{Please describe the shortcomings of the current testing approaches}, with a 60.94\% response rate. 

For those respondents with experience with FMs, we asked two types of questions: ``\textit{Compared with other testing methods, the introduction of LLMs into ADS testing improves the testing capability. (Strong disagree to strong agree)}'' and ``\textit{Compared with other ADS, the introduction of LLMs-based ADS improves the performance. (Strong disagree to strong agree)}''. The rating result is displayed in Figure \ref{fig:rate_LLM}. Overall, both LLMs for testing and LLMs-based ADSs were rated better than ADSs involving VFMs. The average percentage of those who indicated agreement in LLMs-related usage amounted to 65.50\%.

In general, we analyze the demands for both industry and academic researchers separately, and we totally summarized 7 demands: (1) diversity in corner cases, (2) bridging gaps between simulation and real-world testing, (3) more comprehensive testing criteria, (4) effective defenses against current attacks, (5) seamless cross-model collaboration in V2X systems, (6) Test case quality generated by LLMs and (7) reliable cross-modality integration when using FMs. 
 
\subsection{Demand 1: Diversity in corner cases}
Corner cases are extreme or rare driving scenarios under which the ADSs may make wrong decisions due to improper responses, leading to severe traffic accidents \cite{li2022coda}. From the follow-up result, 58.90\% responses have mentioned the lack of diversity of corner cases. Among them, 53.49\% of participants are from industry, and 46.51\% are from research institutes. This demand is primarily put forward by perception module testers and E2E testers, who rely on diverse and realistic edge-case scenarios to evaluate model reliability. Practitioners focus on constructing challenging real-world scenarios, while academia may explore methods for automatically generating or identifying new corner cases. Some responses said that \textit{``Lack of complex corner cases, like a vehicle driving at high speed in heavy fog at night, suddenly encounter a black animal crossing the road.''} and \textit{``The current corner cases are limited, especially when combined with different extreme situations, such as pedestrians suddenly appearing to cross the road in blizzard weather.''} Both of these statements are sophisticated corner cases that are currently lacking, particularly those involving adverse weather conditions, poor visibility, dynamic obstacles, and multi-agent interactions. Therefore, the diversity of corner cases is one of the crucial bottlenecks in ADS testing.

\subsection{Demand 2: Bridging gaps between simulation and real-world testing}
50.68\% of the follow-up responses (including 21.92\% from industry) indicated that simulation environments may fail to fully capture real-world conditions (e.g., lighting variations, sensor noise), potentially causing performance discrepancies when transitioning to real-world deployment. Practitioners are primarily concerned with the discrepancy between simulation and real-world testing, and researchers focus on building higher-fidelity simulation models to bridge the gap between simulation and reality. A key issue raised by many practitioners is the difference between open-loop and closed-loop testing, which affects how ADSs interact with dynamic environments. Closed-loop testing enables real-time feedback, where system outputs influence subsequent inputs, allowing the ADS to dynamically adjust its decisions based on evolving perception data. In contrast, open-loop testing lacks feedback, meaning that perception errors cannot propagate to downstream modules, making it difficult to evaluate the full impact of sensor inaccuracies or decision-making flaws \cite{dona2022virtual}. One response said, ``\textit{The open-loop testing lacks feedback, so it is hard to measure the perception error on downstream modules.}'' 

The professionals argue that both open-loop and closed-loop testing are conducted in simulation environments, but their purposes differ, where open-loop testing is primarily used for evaluating individual components (e.g., perception accuracy), whereas closed-loop testing is crucial for assessing system interactions and decision-making under dynamic conditions. While closed-loop testing is often considered more representative of real-world behavior, most practitioners still rely on simulation-based closed-loop testing before real-world deployment due to the high cost and complexity of physical testing \cite{stocco2022mind}. In this demand, 29.73\% of participants emphasized ``\textit{the lack of dynamic simulation}'', reflecting concerns that existing simulation environments may not fully support closed-loop evaluations. Additionally, some practitioners who worked on control module testing claimed that the dynamics modeling needs to be more precise.

\subsection{Demand 3: More comprehensive testing criteria}
Our follow-up survey results reveal that 30.14\% (with 19.18\% of researchers) of participants identified the lack of comprehensive testing criteria as one of the demands, especially in system-level testing and E2E system testing, which both evaluate the performance of the entire ADS. Researchers are more focused on developing comprehensive evaluation systems, such as new metrics, to improve test coverage and interoperability. Unlike module-level testing, in which specific components can be independently validated, system-level and E2E testing suffer from poor interpretability, making it difficult to pinpoint the root cause of failures. A follow-up response has mentioned that ``\textit{We can acquire outcomes easily when conducting system-level testing, but it is hard to find what part went wrong.}'' This highlights a critical limitation that while system-level testing can indicate ADS performance, it often fails to provide actionable insights into which specific components contributed to errors. 

To address this challenge, researchers have introduced quantitative metrics to evaluate ADSs, such as Vulnerable Road User (VRU), Operational Design Domain (ODD), Conflict Area (CA), etc. \cite{westhofen2023criticality}. However, these metrics do not always correlate with real-world driving quality. One practitioner pointed out this limitation, stating that ``\textit{A good metric does not equal a good driving performance. For example, an ADS that shows a low collision rate in testing as it chooses too conservative a driving strategy, such as braking hard frequently, may lead to rear-end collisions or impact on user experiences.}'' Hence, the comprehensive testing criteria is one of the demands for ADS testing.

\subsection{Demand 4: Effective defenses against current attacks}
26.03\% of follow-up responses came up with concerns about safety by referring to various current attacks, including adversarial, sensor, cyber, and inference attacks. Out of these participants, 84.21\% are conducting academic work, indicating that the industry has fewer concerns about ADS attacks. Practitioners are concerned with deploying defenses in real-world systems, such as adversarial training and encrypted communication. In this demand, 15.79\% of participants claimed the lack of polygonal defenses, and one of them said, ``\textit{Autonomous vehicle security not receive enough attention (maybe only until the accident happens).}'' This reflects a reactive rather than proactive approach to ADS security, suggesting that current defenses are often implemented only after vulnerabilities have been taken advantage of in real incidents, rather than being integrated into the system from the outset. Regarding specific attack vectors, 73.68\% of practitioners emphasized adversarial attacks, which exploit small perturbations in sensor inputs (e.g., images, LiDAR point clouds) to mislead ADS perception models. While adversarial training is a common defense, many practitioners noted that it requires extensive computational resources and can introduce latency, which is often unacceptable in real-time, safety-critical ADS operations. Besides, sensor attacks obtained 52.63\% attention, in which the main focus is jamming (e.g., sending interference signal), spoofing (e.g., sending forging signal), and physical damage attack. 

Moreover, 26.32\% of responses reported that most of the cyber attacks involved the V2X system, ``\textit{In ADSs, some communication paths are unique, and a single point of failure for communications makes the vehicle unable to obtain the real-time data.}'' Additionally, 10.53\% of participants mentioned inference attacks in ADSs, and one of the statements said that ``\textit{In collaborative perception scenarios, such as V2X system, real-time data sharing between vehicles may disclose sensitive information.}'' This reflects the growing challenge of privacy-preserving cooperative ADS perception, where ensuring effective data sharing while mitigating privacy risks is an open research problem. The prevalence of these security concerns highlights the urgent demand for stronger, multi-layered defense mechanisms in ADSs.

\subsection{Demand 5: Seamless cross-model collaboration in V2X systems}
V2X ADSs enhance autonomous driving by enabling vehicles to share real-time data and contribute to traffic efficiency and safety. However, current V2X communication among different vehicles has an inevitable challenge in seamless cross-model collaboration, where autonomous vehicles equipped with different DL models and sensors cannot effectively communicate and make joint decisions due to compatibility issues.

Almost all V2X users in our survey identified model compatibility as a major barrier to the widespread deployment of V2X-enabled ADSs, 83.33\% of them are industry practitioners. The primary concern stems from interoperability challenges between different automobile manufacturers and their proprietary DL models. Specifically, a response said, ``\textit{If an autonomous vehicle is in the V2X DNN of a Tesla company, it is difficult to transmit data with vehicles that joined other automobile companies, such as BYD.}'' Specifically, the core challenge includes adopting distinct DL architectures and frameworks, using different sensor configurations (e.g., variations in LiDAR, camera, and radar setups) that lead to information inconsistencies, and storing and transmitting data in non-standardized formats, all of which prevent seamless cross-company data exchange and real-time decision-making. 

\subsection{Demand 6: Test case quality generated by LLMs}
The advantages of FMs in ADSs include scalability, adaptability, and automation, enabling more diverse scenario construction. VFMs can process and generate realistic driving scenes, while LLMs offer significant advantages in generating test cases in ADSs by reducing manual effort and increasing diversity in testing scenarios. However, LLMs often struggle with real-world consistency, which refers to their ability to align with the physical laws of the world, adhere to traffic regulations, and maintain logical coherence across generated scenarios. This inconsistency leads to test cases that may defy the laws of physics (e.g., vehicles accelerating instantly from 0 to 100 mph), violate established traffic rules (e.g., traffic lights changing unpredictably), or contain contradictions in the sequence of events (e.g., a vehicle stopping at an intersection but still being described as moving forward). 76.19\% of responses put forward the testing scenarios generated by LLMs are not accurate (violation of real-world physical or inconsistent logic), most of them are researchers (81.25\%), and one of the responses declared that ``\textit{Repeated modifications and confirmations with the LLMs are necessary to generate a satisfactory critical scenario,}'' highlighting the inefficiency of the current process. It implies that the LLMs lack a deep understanding of real-world physical interactions, affecting testing scenario generation. Besides, professionals also claimed that the limitation of LLMs in context and multi-modal understanding exacerbates the situation of non-accurate testing scenarios.

\subsection{Demand 7: Reliable cross-modality integration when using FMs}
Apart from test case quality, cross-modality adaptation poses a significant challenge. LLMs, trained on textual data, struggle to process raw sensor inputs (e.g., camera, LiDAR, radar), requiring manual conversion into text, which can lead to errors or missing details. Additionally, translating real-world scenarios into language-based prompts often results in vague test cases, as natural language lacks the precision to capture sensor-driven data. Even with multi-modal FMs, inconsistencies arise in unifying low-level sensor fusion (e.g., camera and LiDAR for object detection) with high-level decision-making (e.g., safe driving maneuvers), creating incoherent tests across modalities.

42.85\% of FMs users mentioned the cross-modality adaptation between NLP and vision. This demand involves frontier topics that are primarily driven by academic research, which extract more attention from researchers. Specifically, FMs in ADS testing may struggle with integrating full-scale awareness (an understanding of all relevant objects and events) in complex driving environments. As one practitioner noted, ``\textit{LLMs are hard to get the input of the scenario, we have to translate the scenario into language,}'' underscoring the fundamental challenge of mapping low-level sensor data into high-level textual descriptions. The inherent ambiguity of natural language further complicates the precise generation of required test scenarios, increasing the risk of misinterpretation in testing. These limitations collectively highlight the effective integration of FMs into ADS testing, especially when bridging perception and reasoning across different modalities.

\section{Literature Review, Current Challenges and Future Direction} \label{literature}
In this section, we summarize the literature related to the demands we identified in Section \ref{demands}, followed by analyzing the challenges and future directions. This work contains 105 references in total, including software engineering journals (TSE, TOSEM, JSS, IST, ESE, SOSP, and SPE) and conferences (ICSE, ISSTA, ASE, FSE, QRS, ICST, and COMPSAC), intelligence vehicle journals (TITS, TII, RAL, IV, and TIV) and conferences (ICRA, CoRL, and DSC), and artificial intelligence conferences (NeurIPS, ICML, AAAI, CVPR, ICCV, and ECCV).

\subsection{Challenges and Directions for Demand 1}

\noindent \textbf{Literature:} We briefly discussed the testing critical scenario approaches in Section \ref{approaches}, including knowledge-based \cite{deng2022declarative, li2023simulation, guo2024semantic}, search-based \cite{tian2022mosat, huai2023sceno, tian2022generating, zhou2023specification, babikian2023concretization, giamattei2024causality, li2024viohawk}, and data-driven \cite{amini2020learning, zohdinasab2024focused, neelofar2024identifying} methods. In the knowledge-based approach, Deng \textit{et~al.} introduced a rule-based metamorphic testing framework for flexible testing scenarios \cite{deng2022declarative}. Li \textit{et~al.} focused on generating test cases for road topology, especially junctions \cite{li2023simulation}. Guo \textit{et~al.} proposed FuzzScene, which generates perturbations to capture rare behaviors \cite{guo2024semantic}.
In the search-based approach, Zhou \textit{et~al.} combined specification-based methods with search techniques to automatically find test cases that violate specifications \cite{zhou2023specification}. The MOSAT uses genetic algorithms to generate test scenarios \cite{tian2022mosat}, and scenoRITA identifies and eliminates duplicate scenarios \cite{huai2023sceno}. Tian \textit{et~al.} leveraged real-world traffic data to generate road topology test scenarios \cite{tian2022generating}.
In the data-driven approach, Amini \textit{et~al.} introduced VISTA, a data-driven simulation to generate test cases for transformations and perturbations \cite{amini2020learning}. Zohdinasab \textit{et~al.} developed DeepAtash-LR, combining data-driven methods with search-based strategies for resource-efficient ADS testing \cite{zohdinasab2024focused}. Additionally, corner case coverage has drawn significant attention. Zhang \textit{et~al.} categorized corner cases and evaluated the suitability of existing metrics and datasets \cite{zhang2022finding}. Li \textit{et~al.} created CODA, a real-world road corner case dataset \cite{li2022coda}, and Biagiola \textit{et~al.} proposed GenBo, which generates challenging scenarios at the behavior boundary \cite{biagiola2024boundary}.

\noindent \textbf{Challenges:} Many papers discuss the limitation of coverage in corner cases \cite{zhang2022finding, song2024industry}. Whether in simulation or real-world testing, generating corner cases is difficult. Although the existing testing datasets cover common driving scenarios, the coverage for extreme scenarios is still an issue. Moreover, the evolving nature of real-world traffic conditions necessitates continuous dataset updates and scenario adaptation, further complicating comprehensive coverage.

\noindent \textbf{Future Directions:} (1) LLMs can generate diverse and rare scenarios by integrating knowledge and complex combinations of objects and environments. Recent studies have demonstrated the potential of LLMs in scenario-based reasoning for autonomous driving \cite{zhang2023automated, guo2024sovar}. However, challenges remain in ensuring scenario validity and physical plausibility. (2) Appeal companies and institute open-source datasets, boost the ADSs community to accumulate more data on corner cases. Efforts such as Waymo Open Dataset \cite{sun2020scalability} have enhanced the diversity of test scenarios. One participant pointed out, ``\textit{To improve ADS robustness, we need more datasets with real-world adversarial scenarios, not just synthetic data.}'' Future work should focus on curating datasets with more adversarial and long-tail driving situations, ensuring better generalization and safety evaluation for ADSs.

\subsection{Challenges and Directions for Demand 2}

\noindent \textbf{Literature:} Simulation platforms like CARLA \cite{dosovitskiy2017carla}, Apollo \cite{apollo2021}, and BeamNG.tech \cite{beamng2022} bridge the gap between real-world and simulation testing by offering realistic environments, diverse scenarios, and automated feedback. Studies have provided insights into their effectiveness \cite{osinski2020simulation, stocco2022mind, biagiola2024two, wang2024drive}. Osinski \textit{et~al.} evaluated simulation-based RL in real-world performance, showing high noise levels with limited randomness \cite{osinski2020simulation}. Stocco \textit{et~al.} examined the transferability between simulation and real-world testing, identifying gaps due to random noise and mechanical stochastic effects \cite{stocco2022mind}. Dai \textit{et~al.} introduced SCTrans to generate simulation scenarios from real-world traffic datasets, but it lacked other objects, which may cause discrepancies \cite{dai2024sctrans}. Wang \textit{et~al.} tested an E2E ADS in the real world, achieving good performance but with slow inference speed \cite{wang2024drive}.

\noindent \textbf{Challenges:} Although simulators can provide a large number of testing data, the uncertainty of the real-world cannot be fully covered, which is mainly attributed to the discrepancies in the received data from sensors between the simulation platforms and the real-world \cite{prakash2021multi, suo2021trafficsim, zhang2022finding, xu2023v2v4real, ding2023survey}. This results in ADSs that perform well in simulators but degrade in the real-world. Additionally, the imprecise dynamic modeling of real-world physics and traffic behavior further exacerbates the problem \cite{haq2020comparing, xiao2022deep, stocco2023model, cheng2024rethinking}. Consequently, achieving reliable transferability between simulated and real-world testing remains an ongoing challenge.

\noindent \textbf{Future Directions:} (1) Testing on hybrid simulation and real-world by screening valuable testing scenarios in simulators and then validating them in real-world experiments, this hybrid approach improves testing efficiency while ensuring scenarios reflect real-world conditions. Simulators can identify valuable scenarios, which are then tested in controlled real-world settings, bridging the gap between simulation limitations and real-world traffic complexities. (2) Explore the fusion technology of multi-modal sensors, especially testing when sensor data fails in extreme environments to enhance the ADSs' robustness. This would improve ADS reliability by ensuring it can adapt and function even when typical sensor inputs fail or are inaccurate. One professional noted, ``\textit{Heavy rain or fog often degrades LiDAR and camera performance, making it critical to rely on radar or sensor fusion techniques.}'' Recent advancements in multi-sensor fusion, particularly in real-world testing \cite{xu2023v2v4real}, can significantly improve ADS robustness in dynamic and unpredictable environments.

\subsection{Challenges and Directions for Demand 3}

\noindent \textbf{Literature:} Evaluation metrics for simulation and real-world testing are designed to measure the safety, reliability, and ability of ADSs, typically including collision rates, trajectory deviations, passenger comfort, etc. In system-level and E2E simulation testing of ADSs, many studies utilized prevalent simulation benchmarks \cite{huang2020multi, wu2022trajectory, coelho2024rlfold} like CoRL2017 \cite{dosovitskiy2017carla}, NoCrash \cite{codevilla2019exploring} and CARLA Leaderboard \cite{carla} to assess performance. The benchmarks focus on criteria such as task success rate, infraction, navigation, etc. CoRL2017 pays more attention to decision-making ability in complex navigation tasks \cite{ishihara2021multi}, NoCrash mainly evaluates the robustness and adaptability of automatic driving systems in complex traffic scenarios \cite{ahmed2021policy}, and CARLA Leaderboard provides a comprehensive assessment framework \cite{haq2023many, lu2024epitester}. Laurent \textit{et~al.} presented an alternative method that combines different parameters for testing to evaluate different decision outputs, thus ensuring each decision path of the ADS is tested \cite{laurent2023parameter}. Besides, in the real-world testing, the criteria also mainly focused on lane-following and obstacle avoidance \cite{wang2024drive}.

\noindent \textbf{Challenges:} Despite the widespread use of simulation benchmarks such as CoRL2017, NoCrash, and CARLA Leaderboard for evaluating ADS performance, the existing evaluation metrics often fail to address the full spectrum of challenges faced by autonomous systems in real-world environments. One of the primary limitations is that these metrics focus on specific, isolated aspects of ADS performance. However, they do not sufficiently assess the long-term stability or robustness of ADSs during continuous operation. This oversight becomes crucial when considering the need for an ADS to operate safely and reliably over extended periods, often under varying conditions \cite{mozaffari2020deep}. Additionally, these metrics lack the ability to provide granular feedback on system failures, making it difficult to diagnose performance issues or bottlenecks. This limitation also hinders the ability to compare ADS performance directly with human driving behavior \cite{zhao2022cadre, biagiola2024two}.

\noindent \textbf{Future Directions:} (1) Consider the long-term performance of ADSs and assess adaptability in multi-tasking scenarios. This includes assessing the system's ability to handle diverse and dynamic environments, particularly in multi-tasking scenarios where the system must handle multiple tasks simultaneously or transition between tasks smoothly. Therefore, one participant emphasized that ``\textit{We need better ways to test how ADSs adapt to long-term driving scenarios instead of just short tests.}'' (2) Introduce comprehensive feedback systems that measure reaction time in complex scenarios and assess responses to unexpected driving scenarios.

\subsection{Challenges and Directions for Demand 4}

\noindent \textbf{Literature:} We investigate attacks frequently mentioned in the responses: physical, cyber, inference, and adversarial attacks. Physical attacks disrupt sensor data or fabricate signals using external hardware. Two common examples are jamming \cite{pham2021survey} and spoofing attacks \cite{tang2023survey}. Jamming adds noise to degrade sensor data, while spoofing injects signals during data collection. Beyond model-agnostic defenses, Sun \textit{et al.} proposed an ADS defense against sensor attacks, though it leads to false alarms \cite{sun2020towards}. Cao \textit{et al.} compared existing defenses, noting that the best defense only reduces the attack success rate to 66\% \cite{cao2021invisible}.

Cyber-attacks on vehicle systems and transport infrastructure pose significant threats to V2X communication \cite{kim2021cybersecurity}. For instance, attackers can control the upload process of HD Maps to the cloud \cite{deng2021deep}. Denial of Service attacks can also disrupt V2X network availability \cite{pethHo2024quantifying}. Defenses against cyber-attacks span areas like data transmission, authentication, and network routing, though they require considerable computational resources \cite{kim2021cybersecurity}. Additionally, inference attacks, evaluated in IoT and cloud techniques, pose threats \cite{gong2022private, yang2023practical}. He \textit{et~al.} showed that membership inference attacks leak information about the perception training dataset, with defenses like Gaussian noise and differential privacy potentially reducing model accuracy \cite{he2020segmentations}.

Adversarial attacks, which involve pixel-level perturbations to input data, cause models to make incorrect decisions \cite{wu2023adversarial, xiang2023v2xp, wang2022adept, von2023deepmaneuver}. ADEPT generates adversarial attacks based on ADS feedback from various test scenarios \cite{wang2022adept}. Von \textit{et~al.} applied adversarial testing to vehicle trajectories to refine perturbations \cite{von2023deepmaneuver}, while Wu \textit{et~al.} developed white-box adversarial attacks against the E2E system \cite{wu2023adversarial}, and Xiang \textit{et~al.} integrated adversarial attacks into the V2X system \cite{xiang2023v2xp}. The most common defense is adversarial training, though it requires significant resources \cite{xiang2023v2xp, von2023deepmaneuver}.

\noindent \textbf{Challenges:} These attacks can affect multiple components of an ADS simultaneously, requiring defense mechanisms that address various dimensions of the system, including data integrity, communication security, sensor fusion, and more \cite{deng2021deep, zhong2022neural, zhang2022finding, tang2023survey}. A major challenge lies in ensuring that defenses can work across the physical, communication, and model layers to provide robust protection without significantly hindering system performance.

\noindent \textbf{Future Directions:} (1) Construct a unified, comprehensive security assessment standard or framework that is flexible to adapt to emerging attack strategies and ensure a holistic defense across different ADS components \cite{zhang2022finding}. The professionals agree that ``\textit{Different companies use different security testing criteria, making it hard to compare ADS security levels objectively.}'' Therefore, a well-defined framework should integrate attack simulation, real-time threat detection, and system-wide robustness evaluation. (2) Develop ADSs that can sense threats in real-time and respond dynamically. This would require the implementation of adaptive security mechanisms that continuously monitor the vehicle’s environment, system state, and communication network for potential threats \cite{tang2023survey}.

\subsection{Challenges and Directions for Demand 5}

\noindent \textbf{Literature:} Communication and cross-model collaboration are inevitable issues in the V2X system \cite{sedar2021standards, martinez2021security, yang2022edge}. Sedar \textit{et~al.} presented an onboard unit that contributes to communication among various equipment manufacturers, who leveraged different V2X protocols \cite{sedar2021standards}. Besides, the standard communication protocols and Ethernet-based networks have leverage widespread \cite{martinez2021security}. However, different organization develops their own DNN models for ADSs, which impacts model compatibility. The common DNN model in ADSs, including Tesla's Full Self-Driving (FSD) DNN \cite{tesla2020}, Apollo DNN \cite{apollo2021}, Waymo DNN \cite{sun2020scalability}, etc. Among them, Tesla's FSD and Waymo DNNs rely on their closed-loop ADS, which makes it difficult to exchange data with other ADSs seamlessly.

\noindent \textbf{Challenges:} The research on model compatibility among multi-vendors is limited \cite{liu2023towards}. The DNN models developed by different automobile manufacturers utilized different data processing methods, feature extraction methods, or differences in decision modules, making distinct ADS difficult to collaborate seamlessly. For example, different V2X interaction models choose different data fusion strategies (early, intermediate, late) according to their application requirements.

\noindent \textbf{Future Directions:} (1) Introduce knowledge distillation to extract useful knowledge from different DNN models and transfer it into a common lightweight model that can be applied across platforms \cite{gou2021knowledge}. One of the professionals states that ``\textit{If we can distill a smaller, cross-model collaboration ADS, it would greatly improve real-time decision-making.}'' It proposes that knowledge distillation may serve as a bridge between proprietary ADS models, ensuring cross-platform compatibility and computational efficiency. (2) Establish standardised interfaces for models in V2X systems, calling for models from various vendors to collaborate through the unified interface. This would reduce the barriers to cross-platform collaboration, enabling better communication and cooperation in V2X systems \cite{yang2022edge}. One participant indicates the potential benefits of a standardized framework, ``\textit{If every company uses its own protocol, cross-vehicle communication will always be a challenge. A standardized model interface would significantly improve vehicle coordination in V2X.}''

\subsection{Challenges and Directions for Demand 6 and Demand 7}
Because Demand 6 and Demand 7 are both related to FMs, we present literature relevant to LLMs and VFMs in the same section and discuss the challenges, respectively.

\noindent \textbf{Literature:} Several studies have applied LLMs to generate testing scenarios for ADSs \cite{zhang2023automated, guo2024sovar, lu2024multimodal, wei2024editable, zhang2024chatscene, lu2024diavio}. Zhang \textit{et~al.} \cite{zhang2023automated} used GPT-3.5 to generate metamorphic relations for metamorphic testing. Guo \textit{et~al.} proposed SoVAR, utilizing GPT-4's ability to extract textual information and generate driving trajectories in the Apollo simulator \cite{guo2024sovar}. Zhang \textit{et~al.} also used GPT-4 to create complex testing scenarios in the CARLA simulator \cite{zhang2024chatscene}. Lu \textit{et~al.} leveraged GPT-4 to diagnose safety violations in simulation testing \cite{lu2024diavio}. In VFMs, SAM \cite{kirillov2023segment} segments objects from images without specific training datasets, while DINO \cite{caron2021emerging} uses self-supervised learning to generate visual representations, enabling the generation of diverse testing scenarios in unseen driving environments.

\noindent \textbf{Challenge of Demand 6:} Improving the generated scenario quality is a demand for introducing LLMs for testing ADSs. The current LLMs do not provide satisfactory results in understanding traffic jargon and non-standard phrase structures in accident reports \cite{guo2024sovar, lu2024multimodal}. Specifically, Song \textit{et~al.} leveraged GPT-4 to identify benign scenarios, but the scenarios lacking surrounding objects were recognized as inconsistent \cite{song2024enhancing}. 

\noindent \textbf{Challenge of Demand 7:} LLMs are based on natural languages. \cite{song2024enhancing, wei2024editable, lu2024diavio} translates domain-specific language into natural language or leverages LLMs setup viewpoint’s position and angle by input natural language. However, the tasks for the generation may not match the expectations due to natural language ambiguity. Errors in LLMs' comprehension of commands also affect task accuracy. Besides, testing standardization and consistency are hard to guarantee, and the different test instructions may generate different scenarios.

\noindent \textbf{Future Directions:} (1) Divide test cases into multiple levels (e.g., basic driving behavior, scene complexity, interaction dynamics), allowing LLMs to generate high-quality test cases by controlling different levels. One of the professionals argues that ``\textit{LLMs generate test cases but often lack structure. If we categorize test cases into different levels, we can systematically control scenario difficulty and complexity.}'' Another professional also agrees, ``\textit{Some test cases are too trivial, while others are unrealistic. If we define structured levels, LLMs can better generate relevant and meaningful cases.}'' (2) Building the translation module, transforming the natural language description into the parameterized scene configuration, which helps to generate more controllable scenes. One practitioner claims the limitations of current test case creation, ``\textit{We often describe a scenario verbally. If we had an automated way to translate test descriptions into simulation-ready parameters, it would significantly streamline testing.}'' It shows a direction for setting up a translation layer that bridges high-level test case descriptions with executable simulation parameters, ensuring that generated scenarios are both realistic and reproducible.

\section{Threats to Validity} \label{section7}
We summarize threats to validity following the criteria from \cite{wohlin2012experimentation}, which are divided into four parts: internal, external, construct, and conclusion validity.

\noindent \textbf{Internal Validity.}
To ensure internal validity and avoid sensitive questions, we avoided questions that might make respondents worried about disclosing company information to encourage truthful responses from participants. During the pilot survey, we adjusted the questions based on feedback until all 10 pilots claimed that the questions were reasonable and did not involve sensitive information. Besides, the anonymity and confidentiality of the data were explained to respondents before our survey to ensure their answers did not reveal any personal or company information.

\noindent \textbf{External Validity.}
Two main external threats may affect the generalizability of our findings: participant sampling bias and discussion bias.

Regarding participant sampling bias: \textit{\textbf{(1)}} our recruitment strategy relied on personal networks and GitHub email crawling. While these methods may introduce sampling bias, potentially favoring academic researchers and open-source ADS developers, we ensured diversity within our personal network by including both industry practitioners and academic contacts. Additionally, we applied strict filtering to remove incomplete, low-quality, or irrelevant responses (e.g., from vehicle scheduling). As a result, our final sample comprises 100 validated ADS testers, with 59\% from academia and 41\% from industry, supporting a balanced representation of perspectives. Nonetheless, we acknowledge that practitioners working on confidential, closed-source ADS projects may still be under-represented. To mitigate this further, we explicitly differentiate academic and industry perspectives throughout our analysis. \textit{\textbf{(2)}} Geographical imbalance may exist. While our participants span eight countries, the majority were recruited through our professional network and are from China (24 participants), the USA (16), Japan (12), and the UK, which together represent major markets and deployment paradigms in ADS development. Although countries like Germany and other regions may be underrepresented, the inclusion of these leading regions, covering both aggressive field deployment (e.g., China, USA) and cautious safety-focused strategies (e.g., Japan, UK), helps ensure that our findings reflect diverse and globally relevant testing practices and challenges.

Regarding the discussion bias, the results may be affected by the initial discussion during the study. Discussions with four professionals involved in ADS testing may have been subjective due to their different experiences and backgrounds, resulting in the lack of representation of the general population. To prevent the survey from focusing only on areas more familiar to the professionals, we conducted literature reviews to summarize the large-scale survey with 107 questions. Besides, we invited 10 experienced candidates as pilot respondents to double-check the comprehensiveness of the survey. The survey was distributed to academic researchers and industry practitioners, and 100 responses were received.

\noindent \textbf{Construct Validity.}
If the survey cannot adequately cover the different test dimensions, may lead to misunderstandings about the overall performance of the ADS testing. To improve construct validity, we comprehensively analyzed ADS testing by covering multiple subareas involved, including V2X, module-level testing, system-level testing, etc. Although most surveys are multiple-choice, we have provided alternative answers to receive more voices. In the follow-up section, we designed open-ended questions for distinct respondents to further investigate their demands.

Our study did not explicitly restrict the scope to a particular Society of Automotive Engineers (SAE) automation level, which defines six levels of driving automation from Level 0 (no automation) to Level 5 (full automation) \cite{sae2021j3016}. Although responses suggest alignment with Level-3 to Level-5 systems based on system descriptions and affiliations, this lack of explicit separation may introduce interpretation variance across participants. Nevertheless, this was intended to ensure broad coverage of ADS testing practices across all automation levels, and many companies evaluate features spanning multiple levels (e.g., Level-3 and Level-4) under unified testing frameworks. We inferred the approximate SAE levels based on participants’ affiliations and response content, where practitioners primarily referenced Level-3 and Level-4 systems, while researchers discussed challenges related to Level-4 and Level-5 autonomy. Therefore, the collected responses still reasonably reflect a wide range of ADS testing practices and challenges.

\noindent \textbf{Conclusion Validity.}
Regarding the conclusion validity of the survey, the first two authors selected 90 complete and high-quality responses separately, with a consistency rate of 97.78\%, and finally obtained a total of 100 responses as the thematic analysis foundation. Additionally, we further surveyed participants who met the follow-up criteria with a response rate of 68.22\%. To capture respondents' thoughts, we designed different open-ended questions based on the participants' ratings to determine the responses' consistency.

\section{Conclusion} \label{section8}
This paper presents a comprehensive survey of testing practices for ADSs, including both modular and E2E systems. It highlights the key demands and challenges faced by both industry practitioners and academic researchers. The survey methodology involved discussions with professionals, a detailed survey of ADS testers and researchers, and follow-up open-ended questions. Seven critical demands were identified, with a particular focus on the diversity of corner cases, testing criteria, potential attacks, and V2X interoperability. Additionally, the increasing use of LLMs for generating test scenarios is noted, alongside demands for improving the quality of test cases through FMs. The paper also provides an in-depth literature review of software engineering research to evaluate progress in addressing these challenges. This work offers actionable insights and future research directions to enhance ADS testing methodologies, ultimately contributing to safer and more reliable autonomous driving systems.


\bibliographystyle{cas-model2-names}
\bibliography{cas-refs}

\end{document}